\documentclass[12pt]{article}

\usepackage{amsmath,amssymb,euscript,array,cite}

\usepackage{graphicx}

\newcommand{\startappendix}{
\setcounter{section}{0}
\renewcommand{\thesection}{\Alph{section}}}
\newcommand{\Appendix}[1]{
\refstepcounter{section}
\begin{flushleft}
{\large\bf Appendix \thesection: #1}
\end{flushleft}}

\makeatletter
 
  \@addtoreset{equation}{section}
 \makeatother

\def\delbarslash{\,\,{\raise.15ex\hbox{/}\mkern-9mu {\bar\partial}}}

\def\R{{\bf R}}

\newcommand{\be}{\begin{equation}}
\newcommand{\ee}{\end{equation}}
\newcommand{\bea}{\begin{eqnarray}}
\newcommand{\eea}{\end{eqnarray}}

\newcommand{\EQ}[1]{\begin{equation} #1 \end{equation}}

\newcommand{\SP}[1]{\begin{equation}\begin{split} #1
\end{split}\end{equation}}

\begin{document}
\newcommand{\nd}[1]{/\hspace{-0.5em} #1}
\begin{titlepage}
\begin{flushright}
{\bf March 2006} \\
SWAT/06/458\\
hep-th/0603075 \\
\end{flushright}
\begin{centering}
\vspace{.2in}
{\large {\bf Double Scaling Limits and Twisted Non-Critical Superstrings}} \\

\vspace{.3in}

Gaetano Bertoldi\\
\vspace{.1 in}
Department of Physics, University of Wales Swansea \\
Singleton Park, Swansea SA2 8PP, UK\\
%
\vspace{.2in}
\vspace{.4in}

{\bf Abstract} 

\end{centering}

We consider double-scaling limits of multicut solutions of 
certain one matrix models that are related to  
Calabi-Yau singularities of type A and the respective 
topological  B model via the Dijkgraaf-Vafa correspondence. 
These double-scaling limits naturally lead to a bosonic string 
with $c  \leq 1$. We argue that this non-critical string is given by the topologically
twisted non-critical superstring background which provides the dual description
of the double-scaled little string theory at the Calabi-Yau singularity. 
The algorithms developed recently to solve a generic multicut 
matrix model by means of the loop equations 
allow to show that the scaling of the higher genus terms 
in the matrix model free energy matches the expected behaviour 
in the topological B-model. 
This result applies to a generic matrix model singularity and the relative double-scaling limit.
We use these techniques to explicitly evaluate the free energy at genus one and genus two.

\end{titlepage}

\section{Introduction}

In \cite{BD}, the large $N$ limit of a class of ${\cal N}=1$ supersymmetric $U(N)$ gauge theories
was studied. The theories are in partially confining phase where an abelian subgroup $\hat{G}$
of the gauge group remains unconfined. The large $N$ spectrum contains the usual weakly 
interacting glueballs, which are neutral under $\hat{G}$, and also baryonic states which are electrically
and magnetically charged with respect to $\hat{G}$, whose mass grows like $N$.
The models studied include the $\beta$-deformation  
of ${\cal N}=4$ Super Yang-Mills and ${\cal N}=1$ SYM coupled to
a single adjoint chiral superfield with a polynomial superpotential. 
At some isolated points in the parameter/moduli space of the models, these baryons can become massless, 
and this causes the $1/N$ expansion to break down. However, it is possible
to define a double-scaling limit in which $N$ goes to infinity and the mass $M_B$ of these
states is kept constant. The crucial feature of this double-scaling limit is that there is
a sector of the Hilbert space of the theory which decouples from the rest and has
finite interactions which are weighted by the double-scaling parameter 
$1/N_{eff} \sim \sqrt{T}/M_B$, where $T$ is the tension of the confining string.
Furthermore, it was proposed in \cite{BD} that the dynamics of this emergent sector
has a dual description given in terms of a non-critical superstring of the type 
introduced in \cite{Kutasov:1990ua}.
This dual formulation has the virtue that the worldsheet theory is exactly solvable and that
the background is free from Ramond-Ramond fluxes.

The exact vacuum structure and F-terms of the ${\cal N}=1$ models with an adjoint chiral field and a 
polynomial superpotential can be analyzed by means of the Dijkgraaf-Vafa matrix model 
correspondence \cite{Dijkgraaf:2002fc,DV3,DVPW}. 
Indeed the proposal of \cite{BD} is based on a careful analysis
of these F-terms. The breakdown of the $1/N$ expansion corresponds to a 
singularity of the matrix model spectral curve and therefore of the dual Calabi-Yau.
The baryon states that become massless correspond to $D3$-branes wrapping shrinking
$3$-cycles in the Calabi-Yau.

In \cite{TLG}, this analysis was extended to a more general class of singularities.
Again, it was found that at these particular points in the moduli space certain states
become massless and that in a suitable double-scaling limit, where the mass
of these states is kept fixed, a particular sector of the theory emerges with interactions 
governed by the double-scaling parameter. There are two novel features 
in these models. First of all, contrary to the cases considered in \cite{BD}, 
there is no supersymmetry enhancement in the double-scaling limit.
This is signalled by the fact that the glueball superpotential does not vanish
in the interacting sector.
In fact, this is also one of the reasons why the dual string background 
is not determined explicitly. 
Secondly, in \cite{TLG}, some or all of the states
that become massless are neutral under the abelian subgroup $\hat{G}$ of the $U(N)$ theory 
which remains unconfined. As a consequence, the presence of these extra massless states 
may not affect the coupling constants of $\hat{G}$ but is captured by the higher genus 
terms of the matrix model free energy as in \cite{Vafaconifold}.
These terms control certain F-term interactions 
of the glueball fields with the graviphoton and gravitational backgrounds \cite{AGNT,BCOV}.

Another important feature that emerges from the analysis of \cite{TLG} 
is that these large $N$ double-scaling limits 
correspond to double-scaling limits of the Dijkgraaf-Vafa matrix model
of the same kind that was considered in the "old matrix model" era
to study $c \leq 1$ systems coupled to two-dimensional gravity \cite{DiFrancesco:1993nw}. 
In particular, in \cite{TLG} it was shown 
that the double-scaling limits have a well-defined genus expansion 
in the sense that the genus $g$ free energy of the matrix model $F_g$ scales like 
$\Delta^{2-2g} \sim M_B^{2-2g}$ \cite{TLG}. 
On the other hand, the singularities and double-scaling limits 
considered in \cite{BD,TLG} generally fall into different universality 
classes from the ones usually considered in the old matrix model. 
It is natural to ask what is the
bosonic non-critical string that corresponds to these matrix model double-scaling limits
and what is the relation between the bosonic non-critical string
and the non-critical superstrings that enter in the dual description 
of the models considered in \cite{BD}. 
The answer to the first question is provided by the Dijkgraaf-Vafa 
correspondence \cite{Dijkgraaf:2002fc,Dijkgraaf:2003xk,ADKMV} 
that states that a generic one matrix model with superpotential $W(\Phi)$ 
is mapped to the topological B model on a 
non-compact Calabi-Yau of the form
\EQ{
uv + y^2 + W'(x)^2  + \text{deformations} = 0
} 
In taking a double-scaling limit, we tune the parameters of
the superpotential and the deformation polynomial so that we are in
the neighbourhood of a particular singularity of the above family of Calabi-Yaus.
For instance in \cite{BD} we are close to an $A_{n-1}$ singularity 
\EQ{
uv + y^2 = x^n - \mu  \  ,
\label{An-1Intro}} 
whereas for the $(2,2p+1)$ bosonic minimal model coupled to 2d gravity
we would have
\EQ{
uv + y^2 + x(x - \epsilon_1)^2 \ldots (x - \epsilon_p)^2 = 0  \  .
}
Therefore, we conclude that the bosonic non-critical string corresponding
to the matrix model double-scaling limit of \cite{BD} is the topological B model at
an $A_{n-1}$ singularity. The case $n=2$ corresponds to the  
conifold singularity.

A check that this is consistent is provided by the fact that 
the scaling of the matrix model free energy $F_g \sim \Delta^{2-2g}$ 
matches exactly the scaling of the topological B model free energy 
\EQ{
F_{top,g}  \sim  \left( \int \Omega \right)^{2-2g} \qquad g > 1
}
where $\Omega$ is the holomorphic $3$-form on the Calabi-Yau \cite{BCOV}.
In fact, the double-scaling parameter $\Delta$ corresponds precisely
to the holomorphic volume of the $3$-cycles that vanish at the singularity. 
This is in turn proportional to $M_B$, the mass of the baryonic states, 
which come from $D3$-branes wrapping the shrinking supersymmetric $3$-cycles.
Furthermore, the fact that in the double-scaling limit $F_g \sim \Delta^{2-2g}$ 
is a general result, it does not depend on the particular class of singularities one considers.
It was derived in \cite{TLG} by means of the recent algorithms to solve 
matrix models based on loop equations \cite{Eynard,ChekhovEynard}.
These allow to consider more general classes of singularities than were previously 
accessible via "old matrix model" techniques.

\begin{figure}
\begin{center}
\includegraphics[scale=0.6]{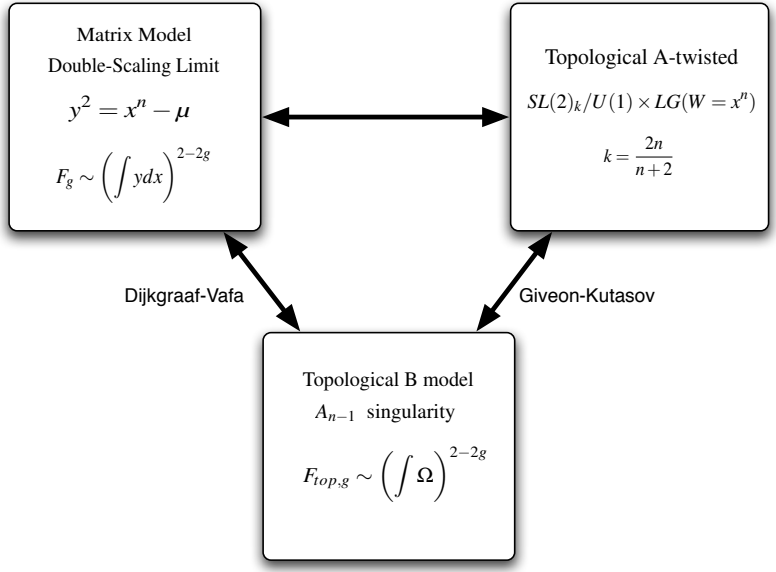} 
\caption{{\small} The bosonic non-critical string defined by the matrix model
double-scaling limit at an $A_{n-1}$ singularity corresponds to the 
$A$-twist of the above $SL(2)/U(1) \times LG$ worldsheet theory.} 
\label{figure1} 
\end{center}
\end{figure}

The non-critical superstring backgrounds that appear as dual 
to the large $N$ double-scaling limits studied in \cite{BD} are of the form
\EQ{
\R^{3,1} \times \left( SL(2)_k/U(1) \times LG(W = X^n)  \right) /  {\bf Z}_n 
\ , \qquad k = \frac{2n}{n+2} \ ,
\label{nonc1}}
where $LG(W)$ denotes a Landau-Ginzburg theory with superpotential $W$.
They were initially introduced in \cite{GK} as holographic duals 
to the 4d double-scaled Little String Theory (DSLST) at a CY singularity of type $A_{n-1}$, 
generalizing the proposal of \cite{GKP} and previous work \cite{LST,holog}.
The non-trivial part of the above background has central charge
\EQ{
\hat{c} = \hat{c}_{sl} + \hat{c}_{LG} = \frac{k+2}{k} + \frac{n-2}{n} = 3  \ ,
} 
and it corresponds to the geometry \cite{OV}
\EQ{
\mu z^{-k} + uv + y^2 + x^n = 0 \  ,
}
where $z,u,v,y,x$ are homogeneous coordinates. 
This is equivalent to \eqref{An-1Intro}.

We argued, using the Dijkgraaf-Vafa correspondence, that the matrix model double-scaling limits considered in \cite{BD} are equivalent to the topological B model at an $A_{n-1}$ singularity \eqref{An-1Intro}.
On the other hand, the matrix model captures the F-terms or topological terms 
of the 4d DSLST \cite{BD} which are given by the topological sector of the 
$SL(2)/U(1) \times LG$ background \eqref{nonc1} \cite{GK}.
Therefore, we expect the non-critical string defined by the matrix model double-scaling limits
to be associated to a topologically twisted $SL(2)_k/U(1) \times LG(X^n)$ background.

This proposal fits nicely with certain known results about the topological twist
of the above background in the conifold case, $n=2$, where the LG model
is trivial. In fact in \cite{GV}, Ghoshal and Vafa argued that the A-twisted 
$N=2$ $SL(2)/U(1)$ supersymmetric coset describes the topological B model 
on a deformed conifold. In \cite{MV}, Mukhi and Vafa had previously shown 
that the above A-twisted 
coset at level $1$ is equivalent to the $c=1$ non-critical bosonic 
string compactified on a circle at self-dual radius.
The open and closed sides of this map were recently analyzed in \cite{Ashok:2005xc}.
Therefore, as a direct generalization of the conifold case, 
we expect that the non-critical bosonic string defined
by the double-scaling limit of \cite{BD} at an $A_{n-1}$ singularity 
should correspond to the $A$-twist of the above 
$SL(2)/U(1) \times LG$ theory. In particular, for $n=2$, the matrix model
double-scaling limit should be equivalent to the $c=1$ string.
This fact can be checked directly on the matrix model side.
Indeed, in the limit, the matrix model spectral curve becomes equivalent to
that of a Gaussian model \cite{TLG} which is equivalent to the $c=1$ non-critical string
\cite{Dijkgraaf:2002fc}.
This particular singularity is obtained from a $2$-cut solution with a 
cubic superpotential in the limit where the two cuts touch each other. 
The fact that this singular limit should be related to the $c=1$ non-critical
string was also observed in \cite{DGKV}.

The relation between strings on non-compact Calabi-Yaus
and non-critical superstring brackgrounds \cite{OV,GK} involving the $N=2$
Kazama-Suzuki $SL(2)/U(1)$ model or its mirror, $N=2$ Liouville theory
\cite{GK,Hori:2001ax,Tong:2003ik}, has been studied by several authors
(see \cite{Eguchi:2004ik,Eguchi:2004yi,Israel:2004jt} and references
therein). 

Furthermore, the relation between the topological sector
of six-dimensional DSLSTs defined at $K3$ singularities, 
the dual topologically twisted non-critical string backgrounds which 
generalize \eqref{nonc1}, and certain non-critical bosonic strings, the 
$(1,n)$ minimal bosonic strings has been recently studied in   
\cite{Rastelli:2005ph,Sahakyan:2005dh,Niarchos:2005ny} (see also 
\cite{NN,Takayanagi0507,Takayanagi0503} for related matters).

In the paper, we are going to use the matrix model double-scaling limit
to study the relative topological B model and non-critical bosonic string.
In section 2, we will review the matrix models studied in \cite{BD,TLG}
and the respective double-scaling limits. In section 3, 
we will review the proof that the genus $g$ matrix model free energy $F_g$
goes like $\Delta^{2-2g}$ as shown in \cite{TLG}. 
As we said, this argument applies to a general matrix model double-scaling limit 
and shows that the scaling of $F_g$ 
is consistent with the expected behaviour of the topological 
B model free energy $F_{g,top} \sim \left( \int \Omega \right)^{2-2g}$ \cite{BCOV}.  
In section 4, we will evaluate the genus one free energy $F_1$ at the $A_{m-1}$ 
singularities considered in \cite{BD}. This gives information on the states that become
massless at the singularity. In section 5, we compute the genus two free energy
relevant to the matrix models considered in \cite{BD}.
The result shows concretely how the double-scaling limit of $F_2$ depends 
on the details of the near-critical spectral curve. In the conifold case, the 
near-critical curve is a Riemann sphere and the general expression 
simplifies drastically and matches the well-known result.

\section{The double-scaling limit}

In this section, we will review the matrix model singularities and relative
double-scaling limits studied in \cite{BD,TLG}.
Consider an ${\cal N}=1$ $U(N)$ theory with a chiral adjoint field $\Phi$ and superpotential
$W(\Phi)$.
The classical vacua of the theory are determined by the stationary points
of $W(\Phi)$
\EQ{
W(\Phi)=N Tr_N \left[ \sum_{i=1}^{\ell+1}\frac{g_i}i \Phi^i\  \right] \  .
\label{bsup}
}
The overall factor $N$ ensures that the superpotential scales 
appropriately in the 't Hooft limit.
For generic values of the couplings, we find $\ell$ stationary points at the zeroes
of 
\EQ{
W'(x) = N \varepsilon \prod_{i=1}^\ell (x-a_i) \  ,  \qquad   \varepsilon \equiv g_{\ell+1}  \  .
} 
The classical vacua correspond to configurations where each of the $N$ eigenvalues of
$\Phi$ takes one of the $\ell$ values, $\{ a_i \}$, for $i=1,\ldots,\ell$.
Thus vacua are related to partitions of $N$ where $N_i \geq 0$ eigenvalues
take the value $a_i$ with $N_1 + N_2 + \ldots N_\ell = N$.
Provided $N_i \geq 2$ for all $i$, the classical low-energy gauge group
in such a vacuum is 
\EQ{
\hat{G}_{cl} = \prod_{i=1}^\ell U(N_i)  \approx  \prod_{i=1}^\ell U(1)_i \times SU(N_i) \  .
}    
Strong-coupling dynamics will produce non-zero gluino condensates
in each non-abelian factor of $\hat{G}_{cl}$. If we define as $W_{\alpha i}$
the chiral field strength of the $SU(N_i)$ vector multiplet in the low-energy theory,
we can define a corresponding low-energy glueball superfield 
$S_i = -(1/32\pi^2) \langle Tr_{N_i} ( W_{\alpha i} W^{\alpha i} ) \rangle$ 
in each factor. Non-perturbative effects generate a superpotential
of the form \cite{CSW,CDSW,VY}
\EQ{
W_{eff}( S_1, \ldots, S_\ell ) =
\sum_{j=1}^\ell N_j ( S_j \log( \Lambda^3_j/S_j) + S_j ) + 2 \pi i \sum_{j=1}^\ell b_j S_j  \  ,
}  
where the $b_j$ are integers defined modulo $N_j$ that label inequivalent 
supersymmetric vacua. 

Dijkgraaf and Vafa argued that the exact superpotential of the theory  
can be determined  by considering a matrix model with potential $W(\hat\Phi)$
\cite{Dijkgraaf:2002fc,DV3} 
\EQ{
\int d \hat \Phi \, \exp \left(  - g_s^{-1} \ Tr \, W(\hat\Phi)   \right)
= \exp \sum_{g=0}^\infty F_g \ g_s^{2g-2}  
}
where $\hat\Phi$ is an $\hat{N} \times \hat{N}$ matrix in the limit $\hat{N} \to \infty$.
The integral has to be understood as a saddle-point expansion around a critical point 
where $\hat{N}_i$ of the eigenvalues sit in the critical point $a_i$. 
Note that $\hat{N}$ is not related to the $N$ from the field theory.
The glueball superfields are identified with the quantities 
\EQ{
S_i = g_s \hat{N}_i \ , \qquad S = \sum_{i=1}^\ell S_i = g_s \hat{N}   
\label{defs}}
in the matrix model and the exact superpotential is 
\EQ{
W_{eff}( S_1, \ldots, S_\ell ) =
\sum_{j=1}^\ell N_j  \frac{\partial F_0}{\partial S_j} + 2 \pi i \sum_{j=1}^\ell b_j S_j  
\label{Wexact}} 
where $F_0$ is the genus zero free energy of the matrix model in the planar limit.
   

The central object in matrix model theory is the resolvent 
\EQ{
\omega(x)=\frac1{\hat N}\text{Tr}\,\frac1{x-\hat\Phi}\ .
}
At leading order in the $1/\hat N$ expansion, $\omega(x)$ is valued on
the spectral curve $\Sigma$, a hyper-elliptic Riemann surface
\EQ{
y^2=\frac1{(N\varepsilon)^2}\big(W'(x)^2+f_{\ell-1}(x)\big)\ .
\label{mmc}
}
The numerical prefactor is chosen for convenience. In terms of this curve
\EQ{
\omega(x)=W'(x)-N\varepsilon y(x)\ .
}
In \eqref{mmc}, $f_{\ell-1}(x)$ is a polynomial of order $\ell-1$
whose $\ell$ coefficients are moduli that are determined by the
$S_i$. In general, the spectral curve can be viewed as a double-cover
of the complex plane connected by $\ell$ cuts. For the saddle-point of
interest only $s$ of the cuts may be opened and so only $s$ of
the moduli $f_{\ell-1}(x)$ can vary. Consequently $y(x)$ has $2s$
branch points and $\ell-s$ zeros:\footnote{Occasionally, for clarity, 
we indicate the order of a polynomial by a subscript.}
\EQ{
\Sigma:\qquad y^2=Z_m(x)^2\sigma_{2s}(x)
\label{mc}
}
where $\ell=m+s$ and 
\EQ{
Z_m(x)=\prod_{j=1}^m(x-z_j)\ ,\qquad
\sigma_{2s}(x)=\prod_{j=1}^{2s}(x-\sigma_j)\ .
}
The remaining moduli are related to the
$s$ parameters $\{S_i\}$ by \eqref{defs}
\EQ{
S_i=g_s\hat N_i=N\varepsilon\oint_{A_i}y\,dx\ ,
}
where the cycle $A_i$ encircles the cut which opens out around the
critical point $a_i$ of $W(x)$. 

Experience with the old matrix model teaches us that double-scaling
limits can exist when the parameters in the potential are varied in
such a way that combinations of branch and double points come
together. In the neighbourhood of such a critical point,\footnote{We
have chosen for convenience to take all the double zeros $\{z_j\}$ 
into the critical region.} 
\EQ{
y^2\longrightarrow C Z_m(x)^2B_n(x)\ ,
\label{redc}
}
where $z_j,b_i\to x_0$, which we can take, without loss of generality, 
to be $x_0=0$. The double-scaling limit involves first taking $a\to0$
\EQ{ 
x=a\tilde x\ ,\qquad z_i=a\tilde z_i\ ,\qquad b_j=a\tilde b_j
\label{dsl}
}
while keeping tilded quantities fixed.
In the limit, we can define the near-critical curve
$\Sigma_-$:\footnote{For polynomials, we use the notation
  $\tilde f(\tilde x)=\prod_i(\tilde
  x-\tilde f_i)$, where $f(x)=\prod_i(x-f_i)$, 
$x=a\tilde x$ and $f_i=a\tilde f_i$.}
\EQ{
\Sigma_-:\qquad y_-^2=\tilde Z_m(\tilde x)^2\tilde B_n(\tilde
x)\ .
\label{ncc}
}
It was shown in \cite{TLG}, generalizing a result of \cite{BD}, that in the limit $a\to0$, in
its sense as a complex manifold, the curve $\Sigma$ factorizes as
$\Sigma_-\cup\Sigma_+$. The complement to the near-critical
curve is of the form
\EQ{
\Sigma_+:\qquad y_+^2=x^{2m+n}F_{2s-n}(x)\ .
}
where $F_{2s-n}(x)$ is regular at $a=0$.

It is important to stress that the above 
singularities are obtained on shell \cite{BD,TLG}.
The family of spectral curves \eqref{redc} corresponds to
vacua of a given field theory. As such, the family satisfies the F-term equations
coming from the exact superpotential \eqref{Wexact} relative to
the model and the choice of semiclassical vacuum.
The solution to the problem of engineering these singularities on shell,
namely the problem of finding a field theory model and 
tree level superpotential whose spectral
curve exhibits the desired singularity in its moduli space,
is explained in detail in \cite{TLG}.  The case where there are 
no double zeroes, $m=0$, has been studied in \cite{Eguchi:2003wv,bert,BD}.
The tree-level superpotential can be taken to be
\EQ{
W(\Phi)=N \varepsilon \ Tr_N \left[\,\Phi^{n+1} - {\cal U}\, \Phi \, \right] \  ,
\label{WAn-1}}
and the relative on-shell spectral curve is 
\EQ{
y^2 = \left( x^n - {\cal U} \right)^2 - {\cal U}_c^2 \ .
\label{curveAn-1}}
At each of the critical values ${\cal U} = \pm {\cal U}_c$, $n$ branch points
collide and the curve has an $A_{n-1}$ singularity. 
For instance, as ${\cal U} \rightarrow {\cal U}_c$
$$
y^2 \approx x^n - ( {\cal U}- {\cal U}_c ) \ .
$$

In the $a\to0$ limit, it was shown in \cite{TLG} 
that the genus $g$ free energy gets a dominant contribution from $\Sigma_-$
of the form
\EQ{
F_g\thicksim \big(Na^{(m+n/2+1)}\big)^{2-2g}\ .
\label{dsp}
}
Note that in this equation $N$ is the one from the field theory and
not the matrix model $\hat N$. This motivates us to define the 
double-scaling limit \cite{BD,TLG}
\begin{equation}
a \rightarrow 0 \ , \qquad N \rightarrow \infty
\ , \qquad \Delta \equiv N a^{m+n/2+1} = \text{const} \ .
\label{limitI1}\end{equation}
Moreover, the most singular terms in $a$ in \eqref{dsp} depend only on the
near-critical curve \eqref{ncc} in a universal way.

Observe that Eq. \eqref{dsp} matches the expected behaviour of the 
topological B model free energy at the singularity \cite{BCOV}.
In fact, as can be seen from \eqref{redc} and \eqref{dsl}
\EQ{
\Delta \sim N \int y\,dx   \  .
\label{olalala2}}
More precisely, the double-scaling parameter is proportional
to the period of the one-form $y\,dx$ on one of the cycles that 
vanish at the singularity.
Moreover, this one-form corresponds to the reduction
of the holomorphic $3$-form $\Omega$ on the underlying Calabi-Yau geometry
\EQ{
uv + y^2 = W'(x)^2 + f(x) 
}
\EQ{
\Omega = \frac{dudvdx}{\sqrt{uv - W'(x)^2 - f(x)}} \ .
}
This comes from the fact that $3$-cycles in the Calabi-Yau correspond
to two-spheres fibered over the complex $x$ plane. 
In particular 
\EQ{
 \int \Omega \sim \int y\, dx 
}
where $\Omega$ is integrated on a vanishing $3$-cycle in the Calabi-Yau
that reduces to one of the vanishing one-cycles in the matrix mode spectral
curve.
Putting everything together, we find that 
\EQ{
F_g \sim \Delta^{2-2g} \sim \left( \int y\,dx \right)^{2-2g} 
\sim \left( \int \Omega \right)^{2-2g} 
}
which is precisely the behaviour we expect for the free energy of the 
topological B model on the Calabi-Yau \cite{BCOV}, in agreement 
with the Dijkgraaf-Vafa correspondence. 
In section \ref{DoubleFg}, we will review the proof of Eq.\eqref{dsp} \cite{TLG} 
using the algorithms of \cite{Eynard,ChekhovEynard} 
based on the matrix model loop equations and we will see relation \eqref{olalala2}
arise naturally.    
It is worth stressing that this result is general, in the sense that is does
not depend on the specific kind of singularities one considers. In this particular respect,
the methods used are more powerful than "old matrix model techniques", where
one is usually limited to considering one-cut matrix model solutions.

\section{The Matrix Model double-scaling limit: higher genus terms}\label{DoubleFg}

In this section, we will be concerned with the behaviour of the 
higher genus terms of the matrix model free energy in
the limit $a\to0$.
The most powerful methods for calculating the $F_g$ involve
orthogonal polynomials (see the reviews \cite{DiFrancesco:1993nw}); 
however, these techniques have only been
successfully applied to the case when the near critical curve has at
most two branch points (but any number of zeros). The only known way
to calculate the $F_g$ in general involves analysing the loop
equations and in particular using the algorithms recently developed 
in \cite{Eynard,ChekhovEynard}.
In the following, we will review these algorithms and the proof 
based on them that
\EQ{
F_g \sim \Delta^{2-2g} 
}
which was given in \cite{TLG}.  As we said above, this is the 
behaviour we expect given the Dijkgraaf-Vafa correspondence
between the matrix model and the topological B model.

\subsection{The loop equations}

The $p$-loop correlator, or $p$-point loop function, is defined as
\SP{
W(x_1,\ldots,x_p) 
&\equiv {\hat N}^{p-2} \Big\langle \, \text{tr} \frac{1}{x_1-\hat\Phi}
\cdots \text{tr} \frac{1}{x_p - \hat\Phi} \, \Big\rangle_\text{conn}\\
&= \frac{d}{dV}(x_p)  \cdots
\frac{d}{dV}(x_1) F \ , \qquad  p\geq 2\ .
}
It has the following genus expansion
\EQ{
W(x_1,\ldots,x_p) = \sum_{g=0}^\infty \frac{1}{\hat N^{2g}}
W^{(g)}(x_1,\ldots,x_p) \ .
}
In \cite{Eynard}, Eynard found a solution to the matrix model loop
equations that allows to write down an expression for these
multiloop correlators at any given genus in terms of a special set of
Feynman diagrams.    
The various quantities involved depend only on the spectral curve of the 
matrix model and in particular one needs to evaluate residues of certain
differentials at the branch points of the spectral curve.

This algorithm and its extension to calculate 
higher genus terms of the matrix model free energy \cite{ChekhovEynard}
represent major progress in the solution
of the matrix model via loop equations \cite{ACKM,Akemann,AmbjornAkemann,Makeenko}. 
This is particularly important because the orthogonal polynomial approach
seems to fail in the multi-cut ($>2$) case.
A nice feature of these algorithms is that they show directly 
how the information is encoded in the spectral curve.
In particular, we will be able to make some precise statements on 
the double-scaling limits of higher genus quantities simply
by studying the double-scaling limit of the spectral curve and its
various differentials.  

Given the matrix model spectral curve for an $s$-cut solution in the
form \eqref{mc} the genus zero $2$-loop function is given by
\SP{
W(x_1,x_2) 
&= -\frac{1}{2(x_1-x_2)^2} +
\frac{\sqrt{\sigma(x_1)}}{2\sqrt{\sigma(x_2)}(x_1-x_2)^2}\\
&-\frac{\sigma'(x_1)}{4(x_1-x_2)\sqrt{\sigma(x_1)}\sqrt{\sigma(x_2)}}
+ \frac{A(x_1,x_2)}{4\sqrt{\sigma(x_1)}\sqrt{\sigma(x_2)}}\ ,  
\label{2loop}
}
where $A$ is a symmetric polynomial given by 
\EQ{
A(x_1,x_2) = \sum_{i=1}^{2s} \frac{ {\cal L}_i(x_2) \sigma(x_1)
}{x_1-\sigma_i}\ , 
\label{A12}
}
where
\EQ{
{\cal L}_i(x_2) = \sum_{l=0}^{s-2} {\cal L}_{i,l}x_2^l =  -
\sum_{j=1}^{s-1} L_j(x_2) \int_{A_j} \frac{dx}{\sqrt{\sigma(x)}}
\frac{1}{(x-\sigma_i)}
\label{LLx}
}
and $s$ is the number of cuts. The polynomials 
$L_j(x)$ are related to the holomorphic 1-forms and defined in Appendix \ref{AppA}
\EQ{
\omega_j = \frac{L_j(x) dx}{\sqrt{\sigma(x)}} \ , \qquad
\int_{A_k} \omega_j = \delta_{jk} \ , \qquad j,k=1,\ldots,s-1 \ .
}

The genus zero $2$-loop function for coincident arguments is
\SP{
W(x_1,x_1)& = \lim_{x_2 \to x_1} W(x_1,x_2) = -\frac{\sigma''(x_1)}{8
\sigma(x_1)} + \frac{\sigma'(x_1)^2}{16 \sigma(x_1)^2} +
\frac{A(x_1,x_1)}{4 \sigma(x_1) }\\
&= \sum_{i=1}^{2s} \frac{1}{16 (x-\sigma_i)^2}
- \frac{\sigma_i''}{16 \sigma'_i(x-\sigma_i)}
+ \frac{{\cal L}_i(x)}{4(x-\sigma_i)}\ .
\label{2loopx1x1}
}
The other important object is the differential
\begin{equation}
dS_{2i-1}(x_1,x_2) = 
dS_{2i}(x_1,x_2)
= \frac{\sqrt{\sigma(x_2)}}{\sqrt{\sigma(x_1)}} \left(
\frac{1}{x_1-x_2} - \frac{L_i(x_1)}{\sqrt{\sigma(x_2)}} -
\sum_{j=1}^{s-1} C_j(x_2) L_j(x_1) \right) dx_1\ ,
\label{dSi}\end{equation} 
where $i=1,\ldots, s$ and 
\begin{equation}
C_j(x_2) = \int_{A_j} \frac{dx}{\sqrt{\sigma(x)} }
\frac{1}{(x-x_2)}\ .
\label{Cj}\end{equation}
A crucial aspect of the one-form (\ref{dSi}) is that it is analytic in 
$x_2$ in the limit $x_2 \to \sigma_{2i-1}$ or $\sigma_{2i}$ \cite{Eynard}
\begin{equation}
\lim_{x_2 \to \sigma_i} \frac{dS_i(x_1,x_2)}{\sqrt{\sigma(x_2)}} =
\frac{1}{\sqrt{\sigma(x_1)}} \left( \frac{1}{x_1-x_2} -
\sum_{j=1}^{s-1}  L_j(x_1) \int_{A_j} \frac{dx}{\sqrt{\sigma(x)}}
\frac{1}{(x-x_2)} \,\, \right) dx_1\ .
\label{dSilimit}\end{equation}
The subtlety is that in the definition of (\ref{Cj}),
the point $x_2$ is taken to be outside the loop surrounding the $j$-th cut,
whereas in (\ref{dSilimit}), $x_2$ is inside the contour.
Note also that
\begin{equation}
A(x_1,x_2) = - \sum_{i=1}^{2s} \left( \sum_{j=1}^{s-1} L_j(x_2)
C_j(\sigma_i) \right) \frac{\sigma(x_1)}{x_1 - \sigma_i}
\label{A12explicit}\end{equation} 
and in particular
\begin{equation}
A(x_1,\sigma_i) = {\cal L}_i(x_1) \sigma'(\sigma_i) \ .
\label{A12explicit2}\end{equation}

The expression of $W^{(g)}(x_1,\ldots,x_p)$ can be found
by evaluating a series of Feynman diagrams of a cubic field 
theory on the spectral curve \cite{Eynard}. To this end,
define the set ${\cal T}^{(g)}_p$ of all
possible graphs with $n$ external legs and with $g$ loops.
They can be described as follows: draw all {\it rooted\/} skeleton 
trees ( trees whose vertices have valence $1,2$ or $3$ ),
with $p+2g-2$ edges. Draw arrows on the edges
from the root towards the leaves. Then draw in all possible ways
$p-1$ external legs and $g$ inner edges with the constraint that
all the vertices of the whole graph have valence three, namely
that are always three and only three edges emanating from any given 
vertex. Each such graph will also have some symmetry factor \cite{Eynard}.

Each diagram in then weighted in the following way.
To each arrowed edge that is part of the skeleton tree going from a vertex
labelled by $x_1$ to a vertex labelled by $x_2$ associate the differential 
$dS(x_1,x_2)$ \eqref{dSi}. 
To each non-arrowed edge associate a genus zero $2$-loop
differential $G(x_1,x_2)=W(x_1,x_2)\,dx_1\,dx_2$
and to each {\it internal\/} vertex labelled by $x_1$ associate the 
factor $(2\varepsilon Ny(x_1) dx_1)^{-1}$.  
For any given tree $T \in {\cal T}_p^{(g)}$, with root $x_1$ 
and leaves $x_j, j=2,\ldots,p$ and with $p+2g-2$ vertices 
labelled by $x'_v, v=1,\ldots,p+2g-2$, so that its inner edges are of the
form $v_1 \to v_2$ and its outer edges are of the form $v \to j$,
we define the weight of the graph as follows
\SP{
{\cal W}(T) &= \frac1{(\varepsilon N)^{p+2g-2}}
\prod_{v=1}^{p+2g-2} \sum_{i_v=1}^{2s} \text{Res}_{x'_v \to b_{i_v}}\,
\frac{1}{2 y(x'_v)dx'_v} 
\prod_{\text{inner edges } v \to w} dS_{i_v}(x'_v,x'_w) \\
&\times\prod_{\text{inner non-arrowed edges }v' \to w'} G_2(x'_{v'},x'_{w'})
\prod_{\text{outer edges }v \to j} G_2(x'_v,x_j)   
\label{weight}
}
In order to find an expression for $F_g$, $g>1$, 
one should consider the same graphs relevant for
$W^{(g)}(x_1)$ and do then the following \cite{ChekhovEynard}:

\noindent
{\bf (i)} Eliminate the first arrowed edge of the skeleton tree.
Labelling the first vertex by $x_1$ and the second vertex by $x_2$,
this amounts to dropping the factor $dS(x_1,x_2)$. \\ 

\noindent
{\bf (ii)} The factor $(2\varepsilon Ny(x_2) dx_2)^{-1}$ has to be dropped 
and  replaced by 
\EQ{
\frac{\int^{x_2}_{q_0} y(s)ds}{y(x_2)dx_2}\ .
}
Note that when evaluating the final residues at $x_2 = \sigma_i$, one needs to expand
the above integral by setting $q_0 = \sigma_i$ \cite{ChekhovEynard}.  
It is also understood that the evaluation of the residues starts from the outer
branches and proceeds towards the root. 
This procedure does not apply for the genus one free energy whose expression
has in any case been found via the loop equations in \cite{ChekhovG=1,KMT,DST}.

We will consider the $a\to0$ limit of each element in \eqref{weight}.  
In this respect, it is
useful to choose a new basis of 1-cycles  $\{\tilde A_i,\tilde B_i\}$,
$i=1,\ldots,s-1$, which is specifically adapted to the
factorization $\Sigma\to\Sigma_-\cup\Sigma_+$. 
The subset of cycles with $i=1,\ldots,[n/2]$  
vanish at the critical point while the cycles $i=[n/2]+1,\ldots,s-1$ 
are the remaining cycles which have zero intersection with
all the vanishing cycles.  
Using the results in Appendix \ref{AppA} for the scaling of $L$, it is
straightforward to argue that for a branch point $b_i$ in the critical
region 
\SP{
&dS_i(x_1,x_2)\longrightarrow
d\tilde{S}_i(\tilde x_1, \tilde x_2) =\\ 
&\frac{\sqrt{\tilde{B}(\tilde{x}_2)}}{\sqrt{\tilde{B}(\tilde{x}_1)}} 
\left(
\frac{1}{\tilde{x}_1-\tilde{x}_2} 
- \frac{\tilde{L}_i(\tilde{x}_1)}{\sqrt{\tilde{B}(\tilde{x}_2)}} -
\sum_{j=1}^{p} 
\tilde{C}_j(\tilde{x}_2) \tilde{L}_j(\tilde{x}_1) \right) d
\tilde{x}_1\ ,
\label{dSiDSL}
}
where $d\tilde{S}_i$ is the analogous differential 
on $\Sigma_-$, Eq.\eqref{ncc}, and $L_j(x)\to a^{n/2-1}\tilde L_j(\tilde x)$ for
$j\leq[n/2]$. 
Conversely, the differentials $dS_i(x_1,x_2)$ where $i$ labels a branch point
of the spectral curve that remains outside of the critical region
give a vanishing contribution in the double-scaling limit. 
Likewise using equations \eqref{2loop}, \eqref{A12}, \eqref{LLx} and 
\eqref{A12explicit} we have
\EQ{
G(x_1,x_2) = W(x_1,x_2) \,dx_1 dx_2 \quad 
\longrightarrow 
\quad \tilde{G}(\tilde{x}_1,\tilde{x}_2) 
= \tilde{W}(\tilde{x}_1, \tilde{x}_2) \,d\tilde{x}_1 d\tilde{x}_2\ ,
}
where $\tilde{G}(\tilde{x}_1,\tilde{x}_2)$ is exactly the $2$-point 
loop correlator on $\Sigma_-$. 

So far we have seen that the double points of the 
near-critical curve do not play a role
in taking the limit of the differentials.
However, this is not the case for the final two elements of the Feynman rules 
\EQ{
y\, dx \longrightarrow\sqrt{C} 
a^{m+n/2+1}\,\, y_-\, d \tilde{x}  
\label{ydxDSL}
}
and
\begin{equation}
\frac{\int_q^x y(s)ds}{y(x)dx} \longrightarrow 
\frac{\int_{\tilde{q}}^{\tilde{x}} 
y_-(\tilde{s})d\tilde{s}}{y_-(\tilde{x})d\tilde{x}} \ .
\label{intyovery}\end{equation}
To summarize: what we have found is that all the relevant quantities
reduce to the analogous quantities on the near-critical curve in the
limit $a\to0$. In particular, being careful with the overall scaling,
the genus $g$ free energy has the limit
\EQ{
F_g\longrightarrow C^{1-g}
\Delta^{2-2g}\, F_g(\Sigma_-) \ . 
\label{FgDSLsigmam}
}
where we have emphasized that $F_g(\Sigma_-)$ 
depends only on $\Sigma_-$. This
is the result advertized in \eqref{dsp} and the property of universality.
Similiarly, the genus $g$ $p$-point loop functions have the limit
\begin{equation}
W_g(x_1,\ldots,x_p) \,dx_1 \cdots dx_p 
\longrightarrow 
C^{1-g-p/2}
\Delta^{2-2g-p} \,\tilde{W}_g(\tilde{x}_1,\ldots,\tilde{x}_p) 
\,d \tilde{x}_1 \cdots d \tilde{x}_p \ .
\label{WgDSL}\end{equation}

\section{The genus one matrix model free energy}\label{genus1}

In this section, we will consider the double-scaling limit
of the genus one free energy $F_1$ in more detail.
This term gives information on the states that become massless
at the singularity \cite{Vafaconifold,OV}. 
The genus one matrix model free energy has been
studied in the context of the Dijkgraaf-Vafa correspondence in \cite{KMT,DST}.
In particular, the authors of \cite{DST} proposed an expression for a
general multicut matrix model solution based on conformal field
theory arguments by Kostov \cite{KostovCFT} and Moore \cite{Moore}. 
This was later proved by Chekhov \cite{ChekhovG=1}
by means of the matrix model loop equations. 
See also \cite{KMT,Vasiliev} 
for an expression of the matrix model genus one 
free energy inspired by the correspondence with the topological B model.
The general expression is given by
\EQ{
F_1 =
- \frac{1}{24} \log
\left(
\prod_{k=1}^{2s} Z( \sigma_k ) \, 
\left(  \prod_{1 \leq i <  j \leq 2s} (\sigma_i - \sigma_j )\right)^4
\, \left( \det_{i,j=1,\ldots,s-1} N_{ij} \right)^{12}
\right)
\label{F1}}
where $s$ is the number of cuts of the matrix model solution and
\begin{equation}
N_{ij} = \int_{A_j} \frac{x^{i-1}}{\sqrt{\sigma(x)}}\,dx
\qquad i,j = 1, \ldots, s-1
\end{equation}
are periods of the holomorphic one-forms $\frac{x^{i-1}dx}{\sqrt{\sigma(x)}}$
on the reduced spectral curve.  
This formula was derived in \cite{ChekhovG=1} 
by considering the genus one $1$-point 
function $W^{(g=1)}(x)$ and explicitly inverting the relation
\EQ{
\frac{d}{d V}(x) F_1 = W^{(g=1)}(x) \ .
\label{Wg1}}
In the previous section, we have seen that in the double-scaling limit
\EQ{
W^{(g=1)}(x) \rightarrow \frac{1}{\Delta} \tilde{W}^{(1)}(\tilde{x})
}
where $\tilde{W}^{(1)}(\tilde{x})$
is the genus one one-point function 
relative to the near-critical spectral curve ${\Sigma}_-$ 
\EQ{
y_-^2(\tilde{x}) =  \tilde{Z}_m(\tilde{x}) \,\tilde{B}_n(\tilde{x}) \ .
}
We can actually absorb the factor of $\Delta$ in the definition of the 
curve itself
\EQ{
y_-^2 =  \Delta^2 \tilde{Z}_m(\tilde{x})\,\tilde{B}_n(\tilde{x})   \ .
}
Thus we obtain  
\EQ{
\frac{d}{d V}(x) F_1 = W^{(g=1)}(x)  \quad \rightarrow \quad
\tilde{W}^{(1)}(\tilde{x}) = \frac{d}{d \tilde{V}}(\tilde{x}) \tilde{F}_1 \ .
\label{limitFg1}}
We can relate the loop insertion operator $\frac{d}{d V}(x)$ to 
$\frac{d}{d \tilde{V}}(\tilde{x})$ as follows. 
Thanks to the identity \cite{ChekhovG=1,ChekhovEynard}
\EQ{
\frac{d\sigma_i }{d V}(x) =  \chi^{(1)}_i(x)  = 
\frac{1}{2 N \varepsilon Z(\sigma_i) \sqrt{\sigma(x)}} \left( 
\frac{1}{x - \sigma_i} + \ldots 
\right) dx
}
we find
$$
\frac{d\sigma_i }{d V}(x)  \rightarrow 
\frac{a}{2 \Delta \,\tilde{Z}_m(\tilde{\sigma}_i)\sqrt{\tilde{B}_n(\tilde{x})}} \left( 
\frac{1}{\tilde{x} - \tilde{\sigma}_i} + \ldots 
\right) d \tilde{x} =  a \tilde{\chi}^{(1)}_i(\tilde{x}) 
=  a \frac{d \tilde{\sigma_i}  }{d \tilde{V}}(\tilde{x})
=  \frac{d \sigma_i  }{d \tilde{V}}(\tilde{x})  \ .
$$
This implies that
\EQ{
\frac{d}{d V}(x) \quad  \rightarrow \quad \frac{d}{d \tilde{V}}(\tilde{x})  \ .
}
Therefore, by \eqref{limitFg1}, we conclude that in the double-scaling limit 
\EQ{
F_1  \quad \rightarrow \quad 
\tilde{F}_1 = - \frac{1}{24} \log \left(
\Delta^{n} \prod_{i=1}^{n} \tilde{Z}(\tilde{\sigma}_i)  
\left(  \prod_{1 \leq i <  j \leq n} 
(\tilde{\sigma}_i - \tilde{\sigma}_j )\right)^4
\, \left( \det_{i,j=1,\ldots,[n/2]} \tilde{N}_{ij} \right)^{12}
\right) \ ,
\label{F1dsl}}
where $\tilde{N}_{ij}$ are periods on the near-critical spectral curve $\Sigma_-$.
This is strictly correct only modulo the addition of a constant, but this plays
no role when one considers general correlators obtained from
$F_1$ like $W^{(1)}(x)$ in \eqref{Wg1}. We also see that
the double-scaled free energy depends in general on the structure of
the near-critical spectral curve. In this respect, observe that the
general expression of $F_1$, Eq.\eqref{F1}, depends on the
basis of $A$-cycles we choose on the spectral curve. Upon a change
of basis, which would correspond physically to an electric-magnetic
duality transformation, $F_1$ changes non-trivially. The expression
\eqref{F1dsl} contains an implicit choice of basis in which the degeneration
of the original spectral curve $\Sigma$ into $\Sigma_+ \cup \Sigma_-$
is made manifest \cite{BD,TLG}. In particular, as in Section \ref{DoubleFg},
we choose a basis such that
$[n/2]$ of the starting $A$-cycles shrink at the singularity and 
reduce to $A$-cycles on the near-critical spectral curve $\Sigma_-$.

In the case of the $A_{n-1}$ singularities studied in \cite{BD}, 
where the near-critical curve is
\EQ{
y^2_- = \tilde{B}_n(\tilde{x})  
}
we find that in the limit $\Delta_{(n)} \to 0$
\EQ{
F_1 \quad \rightarrow \quad - \frac{n}{24} \  \log \Delta_{(n)} \ .
\label{F1An-1}}
In particular, for the conifold singularity,  $n=2$, 
we retrieve the well-known result
\EQ{
F_1 = -\frac{1}{12} \  \log \Delta_{(2)} \  .
\label{F1conifold}}
In the case of an "old matrix model" singularity, where $m$ double zeroes
collide with one branch point of the reduced spectral curve, $\sigma_0$,
and correspondingly the near-critical spectral curve is trivial,  
Eq.\eqref{F1dsl} yields
\EQ{
F_1 \rightarrow - \frac{1}{24} \log \Delta \,\tilde{Z}(\tilde{\sigma}_0)  
\label{oldF1dsl}
}
which is indeed consistent with the result given in \cite{ACKM}.

The divergence of $F_1$ and equivalently of the topological
B model free energy $F_{top,1}$ in the limit $\Delta \rightarrow 0$ indicates
that there are states in the field theory that become massless at the
singularity \cite{Vafaconifold,OV}. Consider type IIB string theory 
compactified on a Calabi-Yau space in the proximity of a conifold singularity.
Vafa argued that \eqref{F1conifold} is consistent 
with the appearance of a single massless 4d ${\cal N}=2$ hypermultiplet 
in the low-energy theory \cite{Vafaconifold}. 
Similarly, type IIB compactified on a Calabi-Yau 
in the proximity of an $A_{n-1}$ singularity yields a 4d ${\cal N}=2$ theory 
close to an Argyres-Douglas point where mutually non-local electric and 
magnetic particles become massless \cite{AD,ARSW,EHIY}.
These extra massless particles also appear in the ${\cal N}=1$ theories
studied via the Dijkgraaf-Vafa matrix model \cite{Eguchi:2003wv,BD,TLG}.
 
Vafa also made a proposal about a general expression for $F_1$
\EQ{
{\cal F}_1 = F_1 + \bar{F_1} = - \frac{1}{12} \sum_{\text{BPS states}} \log m_i^2 
\label{Vafaprop}}
where the sum is over BPS states of the ${\cal N}=2$ 4d theory.
These states can be electrically and magnetically charged and 
come from $D3$-branes wrapping a supersymmetric $3$-cycle $C_i$ 
in the Calabi-Yau. Their mass is given by
\EQ{
m_i^2 = \frac{\int_{C_i} \Omega \cdot \int_{C_i} \bar{\Omega} }
{\int_{CY} \Omega \wedge \bar{\Omega} } \ .
}
This is a generalization of the conifold result \eqref{F1conifold} 
where $m^2 = | \Delta |^2$
and, as stressed in \cite{Vafaconifold}, it might not be the full answer.
For $A_{n-1}$ singularities with $n$ odd, the genus one expression
\eqref{F1An-1}  is indeed not matching the proposal \eqref{Vafaprop}. 
This is probably due to the fact that the states becoming massless are mutually
non-local. It would be interesting to understand better the nature of this result.

\section{Evaluation of the genus $2$ free energy}\label{F2genustwo}

For one-cut matrix model solutions the method of orthogonal polynomials 
allows to evaluate the matrix model free energy at all genera \cite{DiFrancesco:1993nw}.
In the case of multicut solutions, this technique is not generally available.
In order to evaluate higher genus terms in the free energy, we will resort 
to the recently developed algorithms that provide an exact solution to the matrix model loop  
equations \cite{Eynard,ChekhovEynard} that we reviewed in 
Section \ref{DoubleFg}. 
In particular, we will find the expression for the genus two free energy 
in the case where the spectral curve has no double zeroes. This is relevant for
the $A_{n-1}$ singularities considered in \cite{BD}. 

The genus two \eqref{F2} and the genus one \eqref{F1dsl} results show 
that the double-scaled free energy depends in a complex way on the details 
of the near-critical spectral curve.  However, we will see that 
in the simplest case, namely the conifold singularity, this dependence is trivial
and the expressions simplify drastically. This is due to the fact that 
it is possible to choose a basis of A-cycles on the original spectral curve,
before the double-scaling limit, in such a way that the 
the near-critical curve has genus zero, it is a Riemann sphere. 
Thus we recover the known result \cite{BCOV}
\EQ{
u^2 + v^2 + y^2 + x^2 = \mu  \quad  \rightarrow \quad
F_2 = - \frac{1}{240 \mu^2}  \ .
\label{F2conifold}}

\begin{figure}
\begin{center}
\includegraphics[scale=1]{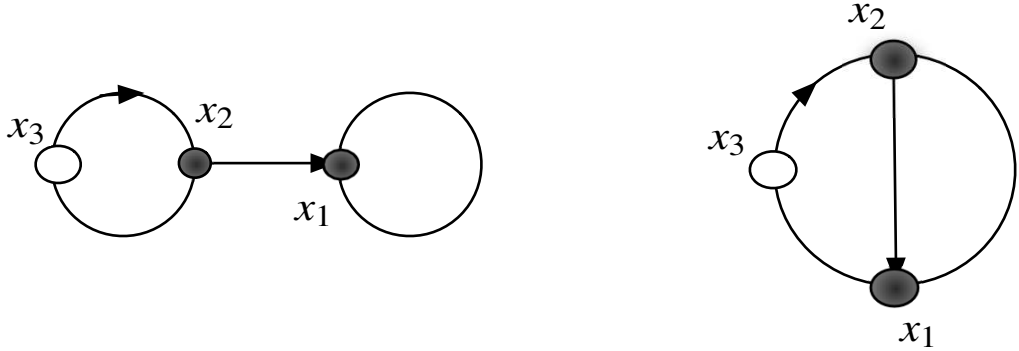} 
\caption{{\small Diagrams $(I)$ and $(II)$.}} 
\label{figure2} 
\end{center}
\end{figure}

\begin{figure}
\begin{center}
\includegraphics[scale=1]{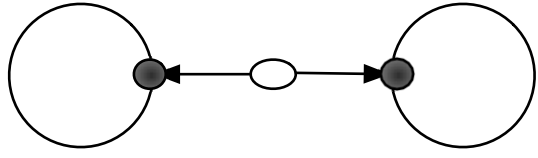} 
\caption{{\small} Diagram $(III)$.} 
\label{figure3} 
\end{center}
\end{figure}

The explicit evaluation of $F_2$ involves calculating three Feynman
diagrams (see Figs.\ref{figure2} and \ref{figure3}).
Diagram $(I)$ is equivalent to
$$
(I) = \int_{C^{x_3} > C^{x_2} > C^{x_1}} \frac{dx_3}{2 \pi i} \frac{dx_2}{2
\pi i}
\frac{dx_1}{2 \pi i}
\frac{\int_{q_0}^{x_3} y(s)ds}{y(x_3) dx_3} 
\frac{dS(x_3,x_2)}{2 y(x_2)}\, W(x_3,x_2)\, \frac{dS(x_2,x_1)}{2 y(x_1)} W(x_1,x_1)
$$
\EQ{
= \int_{C^{x_3} > C^{x_2}} \frac{dx_3}{2 \pi i} \frac{dx_2}{2
\pi i} \frac{\int_{q_0}^{x_3} y(s)ds}{y(x_3) dx_3} 
\frac{dS(x_3,x_2)}{2 y(x_2)}\, W(x_3,x_2)\, W^{(1)}(x_2) \ .
\label{DiagI}
}
Diagram $(II)$ is 
$$
(II) = \int_{C^{x_3} > C^{x_2} > C^{x_1}}
\frac{dx_3}{2 \pi i} \frac{dx_2}{2\pi i} \frac{dx_1}{2 \pi i}
\frac{\int_{q_0}^{x_3} y(s)ds}{y(x_3) dx_3}
\frac{dS(x_3,x_2)}{2 y(x_2)} \frac{dS(x_2,x_1)}{2 y(x_1)}
W(x_3,x_1) W(x_2,x_1)
$$
\begin{equation}
= \int_{C^{x_3} > C^{x_2}} \frac{dx_3}{2 \pi i} \frac{dx_2}{2
\pi i} \frac{\int_{q_0}^{x_3} y(s)ds}{y(x_3) dx_3}
\frac{dS(x_3,x_2)}{2 y(x_2)}\, W(x_3,x_2,x_2)\, \ .
\label{DiagII}\end{equation}

\noindent
Similarly, diagram $(III)$ gives
$$
(III) = \int_{C^{x_3}} \frac{dx_3}{2 \pi i} \frac{\int_{q_0}^{x_3}
y(s)ds}{y(x_3) dx_3} 
\int_{C^{x_1}}  \frac{dx_1}{2\pi i} \frac{dS(x_3,x_1)}{2 y(x_1)} W(x_1,x_1)
\int_{C^{x_2}}  \frac{dx_2}{2\pi i} \frac{dS(x_3,x_2)}{2 y(x_2)} W(x_2,x_2)
$$
\begin{equation}
= \int_{C^{x_3}} \frac{dx_3}{2 \pi i} \frac{\int_{q_0}^{x_3}
y(s)ds}{y(x_3) dx_3} W^{(1)}(x_3)\, W^{(1)}(x_3)\,  \  .
\label{DiagIII}\end{equation}
Finally, as shown in \cite{ChekhovEynard}
\begin{equation}
2 F_2 = 2 (I) + 2 (II) + (III) \  .
\end{equation}
In the following, we will only consider cases where
the spectral curve has no double points, and 
we will set 
\EQ{
y^2 = \varepsilon^2 \sigma_{2s}(x) \ .
\label{notationF2sp}}

{\bf (I)}: Using the expression of $W^{(1)}(x)$ evaluated in the appendix 
\eqref{W(1)} and the differentials $\chi^{(n)}_i(x_3)$, \eqref{chin}\eqref{chini}, we find 
$$
\int_{C^{x_2}} \frac{dx_2}{2\pi i}
\frac{dS(x_3,x_2)}{2 y(x_2)} \, W(x_3,x_2) W^{(1)}(x_2)
= \sum_{i=1}^{2s} \frac{1}{2\varepsilon} \left[ \frac{1}{16 \sigma'_i}
G(x_3,\sigma_i) \chi^{(3)}_i(x_3)
\right.
$$
$$
\left.
+ \left( \frac{1}{16 \sigma'_i} \frac{\partial}{\partial x_2} G(x_3,x_2)|_{x_2=\sigma_i} 
- \frac{\sigma_i''}{32 \sigma_i'{}^2} G(x_3,\sigma_i)
\right) \chi^{(2)}_i(x_3)
\right.
$$
$$
\left.
+  \left( \frac{3 \sigma_i''{}^2-2\sigma_i' \sigma_i'''}{192 \sigma_i'{}^3} G(x_3,\sigma_i)
+ \frac{1}{32 \sigma'_i} \frac{\partial^2}{\partial x_2^2} G(x_3,x_2)|_{x_2=\sigma_i} 
- \frac{\sigma_i''}{32 \sigma'_i{}^2} \frac{\partial}{\partial x_2} G(x_3,x_2)|_{x_2=\sigma_i} 
\right) \chi^{(1)}_i(x_3)
\right.
$$
$$
\left.
+  G(x_3,\sigma_i)   \frac{{\cal L}^{(2)}_i(\sigma_i)}{16\sigma_i'} \chi^{(1)}_i(x_3) 
+ \sum_{j\ne i} \left(
\frac{1}{16} \hat{\chi}^{(2)}_j(\sigma_i) + B_j \hat{\chi}_j^{(1)}(\sigma_i) 
\right) \frac{G(x_3,\sigma_i)}{\sigma'_i}  \chi^{(1)}_i(x_3)  
\right.
$$
\EQ{
\left.
+ \frac{B_i}{\sigma'_i}
G(x_3,\sigma_i) \chi^{(2)}_i(x_3)
+ \left( \frac{B_i}{\sigma'_i} \frac{\partial}{\partial x_2} G(x_3,x_2)|_{x_2=\sigma_i} 
- \frac{B_i \sigma_i''}{2 \sigma_i'{}^2} G(x_3,\sigma_i)
+ B_i \frac{{\cal L}_i(\sigma_i)}{\sigma_i'} G(x_3,\sigma_i) 
\right) \chi^{(1)}_i(x_3)
\right]
\label{AttackI}}
where we introduced 
\EQ{
G(x_3,x_2) = \frac{\sqrt{\sigma(x_3)}}{2(x_3-x_2)^2}
- \frac{\sigma'(x_3)}{4(x_3-x_2)\sqrt{\sigma(x_3)}}
+ \frac{A(x_3,x_2)}{4 \sqrt{\sigma(x_3)}}  \  ,
\label{Gx3x2}}
\EQ{
A(x_1,x_2) = \sum_{i=1}^{2s} \frac{ {\cal L}_i(x_2) \sigma(x_1)
}{x_1-\sigma_i}\ , 
\label{A12cippo}}
\begin{equation}
{\cal L}_i^{(2)}(x_2) =  - \sum_{j=1}^{s-1} L_j(x_2) \int_{A_j}
\frac{dx}{\sqrt{\sigma(x)}} \frac{1}{(x-\sigma_i)^2} \,   .
\label{Bi2}\end{equation}

\noindent
Before we illustrate how to perform the final integration,
let us introduce the following notation
\EQ{
\chi^{(n)}_k(x) = \frac{1}{2 \varepsilon \sqrt{\sigma(x)}}
\,\hat\chi^{(n)}_k(x) \ ,
\quad 
\hat\chi^{(n)}_k(x) = \left( \frac{1}{(x-\sigma_k)^n}
+ {\cal L}^{(n)}_k(x)\right) \ ,
\label{hats1}}
\EQ{
G(x_3,\sigma_k) = \frac{1}{\sqrt{\sigma(x_3)}} \hat{G}(x_3,\sigma_k) \ .
\label{Ghat}}
We find that 
$$
(I) = \frac{1}{4 \varepsilon^2} \sum_{k=1}^{2s} 
\int_{C^{x_3}} \frac{dx_3}{2 \pi i} 
\frac{\int_{A_k}^{x_3} \sqrt{\sigma}(s)ds}{\sigma(x_3)^{3/2}} 
\sum_{i=1}^{2s}  \left[ \frac{1}{16 \sigma'_i}
\hat{G}(x_3,\sigma_i) \hat\chi^{(3)}_i(x_3)
\right.
$$
$$
\left.
+ \left( \frac{1}{16 \sigma'_i} \frac{\partial}{\partial x_2} \hat{G}(x_3,x_2)|_{x_2=\sigma_i} 
- \frac{\sigma_i''}{32 \sigma_i'{}^2} \hat{G}(x_3,\sigma_i)
\right) \hat{\chi}^{(2)}_i(x_3)
\right.
$$
$$
\left.
+  \left( \frac{3 \sigma_i''{}^2-2\sigma_i' \sigma_i'''}{192 \sigma_i'{}^3} \hat{G}(x_3,\sigma_i)
+ \frac{1}{32 \sigma'_i} \frac{\partial^2}{\partial x_2^2} \hat{G}(x_3,x_2)|_{x_2=\sigma_i} 
- \frac{\sigma_i''}{32 \sigma'_i{}^2} \frac{\partial}{\partial x_2} \hat{G}(x_3,x_2)|_{x_2=\sigma_i} 
\right) \hat{\chi}^{(1)}_i(x_3)
\right.
$$
$$
\left.
+  \hat{G}(x_3,\sigma_i)   \frac{{\cal L}^{(2)}_i(\sigma_i)}{16\sigma_i'} \hat{\chi}^{(1)}_i(x_3) 
+ \sum_{j\ne i} \left(
\frac{1}{16} \hat{\chi}^{(2)}_j(\sigma_i) + B_j \hat{\chi}_j^{(1)}(\sigma_i) 
\right) \frac{\hat{G}(x_3,\sigma_i)}{\sigma'_i}  \hat{\chi}^{(1)}_i(x_3)  
\right.
$$
\EQ{
\left.
+ \frac{B_i}{\sigma'_i}
\hat{G}(x_3,\sigma_i) \hat{\chi}^{(2)}_i(x_3)
+ \left( \frac{B_i}{\sigma'_i} \frac{\partial}{\partial x_2} \hat{G}(x_3,x_2)|_{x_2=\sigma_i} 
- \frac{B_i \sigma_i''}{2 \sigma_i'{}^2} \hat{G}(x_3,\sigma_i)
+ B_i \frac{{\cal L}_i(\sigma_i)}{\sigma_i'} \hat{G}(x_3,\sigma_i) 
\right) \hat{\chi}^{(1)}_i(x_3)
\right]
\label{AttackIbis}}

\noindent
The next step is to expand
$$
\frac{\int^{x_3}_{\sigma_k} \sqrt{\sigma(s)}ds}{\sigma(x_3)^{3/2}}
$$
in the proximity of the branch point $\sigma_k$ itself \cite{ChekhovEynard}.
Setting $\epsilon = x-\sigma_k$, we find
$$
\frac{\int_{A_k}^{x_3} \sqrt{\sigma(s)}ds}{\sigma(x_3)^{3/2}} =
\frac{2}{3 \sigma'_k} \left( \, 1 + c_{1,k} \epsilon + c_{2,k} \epsilon^2 +
c_{3,k} \epsilon^3 + {\cal O}(\epsilon^4) \,\right)
$$
where
$$
c_{1,k} = -\frac{3 \sigma''_k}{5 \sigma'_k}
\qquad
c_{2,k} = \frac{3(8 \sigma''_k{}^2 - 5 \sigma'_k \sigma'''_k)}{70 \sigma'_k{}^2}
\qquad
c_{3,k} = - \frac{60 \sigma''_k{}^3 - 76 \sigma'_k \sigma''_k \sigma'''_k + 105 \sigma'_k{}^2 \sigma''''_k}{315 \sigma'_k{}^3}
$$
Finally
$$
(I) =  \sum_{k=1}^{2s}  \frac{1}{6 \varepsilon^2 \sigma'_k} 
\left[ 
\frac{1}{16 \sigma_k'} \left( \frac{c_{3,k} \sigma'_k}{4} + \frac{c_{2,k} A(\sigma_k,\sigma_k)}{4} +
\frac{c_{1,k} (6A^{(1,0)}(\sigma_k,\sigma_k)-\sigma_k''' )}{24}
\right.
\right.
$$
$$
\left.
\left.
+  \frac{6A^{(2,0)}(\sigma_k,\sigma_k) + 12 \sigma_k' {\cal L}^{(3)}_k(\sigma_k) - \sigma_k''''}{48} \right)
+ \frac{1}{16 \sigma_k'}  
\left(
\frac{3 c_{3,k} \sigma'_k + c_{2,k} \sigma_k''}{4}
\right.
\right.
$$
$$
\left.\left.
+ \frac{c_{1,k} (6A^{(0,1)}(\sigma_k,\sigma_k)+\sigma_k'''  + 18 \sigma_k' {\cal L}^{(2)}_k(\sigma_k))}{24}
+ \frac{A^{(1,1)}(\sigma_k,\sigma_k)}{4} 
+ \frac{\sigma_k'' {\cal L}^{(2)}_k(\sigma_k) + 3 \sigma'_k {\cal L}^{(2)}(\sigma_k){}'}{4}
\right)
\right.
$$
$$
\left. 
+  \left( \frac{B_k}{\sigma'_k}  - \frac{\sigma_k''}{32 \sigma_k'{}^2} \right) 
\left(  \frac{c_{2,k}\sigma_k'}{4} + \frac{c_{1,k}A(\sigma_k,\sigma_k)}{4} 
+  \frac{6A^{(1,0)}(\sigma_k,\sigma_k) - \sigma_k'''  + 6 \sigma_k' {\cal L}^{(2)}_k(\sigma_k)}{24} \right)
\right.
$$
$$
\left.
+ \left( \frac{B_k}{\sigma'_k}  - \frac{\sigma_k''}{32 \sigma_k'{}^2} \right) 
\left(  \frac{3 c_{2,k}\sigma_k'}{4} + c_{1,k} \frac{\sigma_k'' +  3 \sigma'_k {\cal L}_k(\sigma_k)}{4} 
+  \frac{6A^{(0,1)}(\sigma_k,\sigma_k) + \sigma_k'''  +  6 \sigma_k'' {\cal L}_k(\sigma_k) + 18 \sigma_k' {\cal L}_k'(\sigma_k)}{24} \right)
\right.
$$
$$
\left. 
+ \frac{1}{32 \sigma'_k} \left(
\frac{5 c_{3,k} \sigma'_k}{2}  + c_{2,k} \frac{2 \sigma_k'' + 5 A(\sigma_k,\sigma_k)}{2}
+ c_{1,k} \frac{\sigma_k''' + 4 \sigma_k'' {\cal L}_k(\sigma_k) +10  \sigma_k' {\cal L}_k'(\sigma_k)  }{4}
\right.
\right.
$$
$$
\left.\left.
+ \frac{6A^{(0,2)}(\sigma_k,\sigma_k) + \sigma_k''''  + 6 \sigma_k''' {\cal L}_k(\sigma_k) + 24 \sigma_k'' {\cal L}_k(\sigma_k)'
+ 30 \sigma_k' {\cal L}''_k(\sigma_k)}{24} 
\right)
\right.
$$
$$
\left.
+ \left( \frac{3 \sigma_k''{}^2-2\sigma_k' \sigma_k'''}{192 \sigma_k'{}^3} 
- \frac{B_k \sigma_k''}{2 \sigma_k'{}^2} 
+ B_k \frac{{\cal L}_k(\sigma_k)}{\sigma'_k}
+ \frac{{\cal L}_k^{(2)}(\sigma_k)}{16 \sigma'_k}
+ \sum_{j \ne k} \frac{1}{\sigma'_k} \left(  \frac{1}{16} \hat{\chi}^{(2)}_j(\sigma_k) + B_j \hat{\chi}^{(1)}_j(\sigma_k)  \right)
\right) 
\right.
$$
\EQ{
\left.
\times
\left(
\frac{c_{1,k} \sigma'_k}{4} +  \frac{A(\sigma_k,\sigma_k)}{2}
\right)
\right]
\label{Ifinal}
}

\noindent
{\bf(II)}: Using the expression of $W(x_2,x_2,x_3)$, Eq.(\ref{W3}), we find
$$
\int_{C^{x_2}} \frac{dx_2}{2\pi i}
\frac{dS(x_3,x_2)}{2 y(x_2)} \, W(x_3,x_2,x_2)
$$
$$
= \sum_{i=1}^{2s} 
\left[ \frac{1}{16} \chi^{(1)}_i(x_3) \chi^{(3)}_i(x_3)
+ \left( \frac{A(\sigma_i,\sigma_i)}{8 \sigma'_i}   
- \frac{\sigma''_i}{32 \sigma'_i}
\right) \chi^{(1)}_i(x_3)\chi^{(2)}_i(x_3)
\right.
$$
$$
\left. +
\left( \frac{A^{(1,0)}(\sigma_i,\sigma_i)}{8\sigma'_i}
+ \frac{A(\sigma_i,\sigma_i)^2}{16 \sigma'_i{}^2}
- \frac{\sigma'_i}{16} \sum_{j \ne i} \frac{1}{\sigma'_j (\sigma_i-\sigma_j)^2}
- \frac{\sigma_i'' A(\sigma_i,\sigma_i)}{16 \sigma'_i{}^2}
\right) \chi^{(1)}_i(x_3)^2
\right.
$$
\begin{equation}
\left.
+ \sum_{j \ne i} \left( 
\frac{1}{16(\sigma_j-\sigma_i)^2} 
\left( \frac{\sigma'_i{}^2 + \sigma'_j{}^2}{\sigma'_i \sigma'_j} \right) 
+ \frac{A(\sigma_i,\sigma_j)^2}{16 \sigma'_i \sigma'_j}
+ \frac{A(\sigma_i,\sigma_j)(\sigma'_j-\sigma'_i)}{16(\sigma_i-\sigma_j)\sigma'_i \sigma'_j} 
\right) \chi^{(1)}_j(x_3) \chi^{(1)}_i(x_3)
\right]
\end{equation}

\noindent
Similarly for $(II)$ we find
$$
(II) = \sum_{k=1}^{2s} \frac{1}{6 \varepsilon^2 \sigma_k'}
\left[ \frac{c_{3,k}}{16} + \frac{{\cal L}^{(3)}_k(\sigma_k)}{16}
+ \frac{c_{2,k}}{16 \sigma'_k} A(\sigma_k,\sigma_k) 
+ \frac{c_{1,k}}{16 \sigma'_k} A^{(1,0)}(\sigma_k,\sigma_k) 
+ \frac{1}{32 \sigma'_k} A^{(2.0)}(\sigma_k,\sigma_k) 
\right.
$$
$$
\left.
+ \left( \frac{A(\sigma_k,\sigma_k)}{8 \sigma'_k} - \frac{\sigma_k''}{32\sigma'_k} \right)
\left( c_{2,k} + {\cal L}^{(2)}_k(\sigma_k) + \frac{c_{1,k} A(\sigma_k,\sigma_k) + A^{(1,0)}(\sigma_k,\sigma_k)}
{\sigma'_k} \right)
\right.
$$
$$
\left. +
\left( \frac{A^{(1,0)}(\sigma_k,\sigma_k)}{8\sigma'_k}
+ \frac{A(\sigma_k,\sigma_k)^2}{16 \sigma'_k{}^2}
- \frac{\sigma'_k}{16} \sum_{j \ne k} \frac{1}{\sigma'_j (\sigma_k-\sigma_j)^2}
- \frac{\sigma_k'' A(\sigma_k,\sigma_k)}{16 \sigma'_k{}^2}
\right)
\left(
c_{1,k} + \frac{2 A(\sigma_k,\sigma_k)}{\sigma'_k}
\right)
\right.
$$
\EQ{
\left.
+ \sum_{j \ne k} 2 \left( 
\frac{1}{16(\sigma_j-\sigma_k)^2} 
\left( \frac{\sigma'_k{}^2 + \sigma'_j{}^2}{\sigma'_k \sigma'_j} \right) 
+ \frac{A(\sigma_k,\sigma_j)^2}{16 \sigma'_k \sigma'_j}
+ \frac{A(\sigma_k,\sigma_j)(\sigma'_j-\sigma'_k)}{16(\sigma_k-\sigma_j)\sigma'_k \sigma'_j} 
\right) \left( \frac{1}{\sigma_k-\sigma_j} +
\frac{A(\sigma_k,\sigma_j)}{\sigma'_j}  \right)
\right]
\label{IIfinal}}

\noindent
{\bf (III)}: From Eqs.(\ref{DiagIII}) and (\ref{W(1)}), we find
$$
(III) = \sum_{k=1}^{2s} \frac{1}{6\varepsilon^2 \sigma'_k}
\left[ \frac{c_{3,k} + 2 c_{1,k} {\cal L}^{(2)}_{k}(\sigma_k) + 2 {\cal L}^{(2)}_{k}{}'(\sigma_k) }{256} 
\right.
$$
$$
\left.
+ \frac{c_{1,k}}{128} \left( \sum_{j\ne k} \frac{1}{(\sigma_k-\sigma_j)^2} + {\cal L}^{(2)}_j(\sigma_k) \right) 
+ \frac{1}{128} \left( \sum_{j\ne k} -\frac{2}{(\sigma_k-\sigma_j)^3} + {\cal L}^{(2)}{}'_j(\sigma_k) \right) + \right.
$$
$$
\left.
+ \frac{B_k}{8} \left( c_{2,k} + {\cal L}^{(2)}_k(\sigma_k)
+ \frac{c_{1,k} A(\sigma_k,\sigma_k) + A^{(1,0)}(\sigma_k,\sigma_k)}{\sigma'_k} \right)
\right.
$$
$$
\left.
+ \sum_{j\ne k} B_j \left( \frac{c_{1,k}}{8} 
\left( \frac{1}{\sigma_k-\sigma_j}  + \frac{A(\sigma_j,\sigma_k)}{\sigma'_j} \right)
+ \frac{1}{8} \left( -\frac{1}{(\sigma_k-\sigma_j)^2} + \frac{A^{(1,0)}(\sigma_k,\sigma_j)}{\sigma'_j}  
\right) \right)
\right.
$$
\EQ{
\left. 
+ \frac{B_k}{8} 
\sum_{j\ne k} \left(  \frac{1}{(\sigma_k-\sigma_j)^2} + {\cal L}^{(2)}_j(\sigma_k)  \right)
+ B_k^2 \left( c_{1,k} + 2 \frac{A(\sigma_k,\sigma_k)}{\sigma'_k} \right)
+ 2 B_k \sum_{j\ne k} B_j \left(  \frac{1}{\sigma_k-\sigma_j} + \frac{A(\sigma_k,\sigma_j)}{\sigma'_j}  \right)
\right]  \ .
\label{IIIfinal}}

\noindent
Therefore, the final expression for the genus two free energy is
$$
F_2 =  \frac{1}{\varepsilon^2} \sum_{i=1}^{2s}  
\left( -\frac{157 {\sigma''_i}^3}{15360 {\sigma'_i}^4}
+\frac{491 {\sigma_i''} {\sigma'''_i}}{46080 {\sigma'_i}^3}
-\frac{35 \sigma_i''''}{3072 {\sigma'_i}^2}
+\frac{35 {\sigma_i''}^2 A(\sigma_i,\sigma_i)}{768 {\sigma_i'}^4}
-\frac{11 \sigma'''_i A(\sigma_i,\sigma_i)}{576 {\sigma_i'}^3}
-\frac{49 {\sigma''_i} A(\sigma_i,\sigma_i)^2}{640 {\sigma_i}^4}
\right.
$$
$$
\left.
+\frac{5 A(\sigma_i,\sigma_i)^3}{96{\sigma'_i}^4}
-\frac{37 {\sigma''_i} {\cal L}^{(2)}(\sigma_i)}{2560 {\sigma_i'}^2}
+\frac{A(\sigma_i,\sigma_i) {\cal L}^{(2)}_i(\sigma_i)}{24 {\sigma_i'}^2}
+\frac{13 {\cal L}^{(2)}{}'(\sigma_i)}{1536 {\sigma'_i}}
+\frac{5 {\cal L}^{(3)}_i(\sigma_i)}{384 \sigma_i'}
-\frac{11 \sigma_i'' {\cal L}_i'(\sigma_i)}{768 \sigma'_i{}^2}
\right.
$$
$$
\left.
+ \frac{A(\sigma_i,\sigma_i) {\cal L}'_i(\sigma_i)}{32 {\sigma'_i}^2}
+ \frac{5 {\cal L}''_i(\sigma_i)}{768 {\sigma_i'}}
- \frac{47 \sigma''_i A^{(0,1)}(\sigma_i,\sigma_i)}{7680 {\sigma'_i}^3}
+ \frac{5 A(\sigma_i,\sigma_i) A^{(0,1)}(\sigma_i,\sigma_i)}{384 {\sigma'_i}^3}
+ \frac{A^{(0,2)}(\sigma_i,\sigma_i)}{768 {\sigma'_i}^2}
\right.
$$
$$
\left.
-\frac{89 \sigma''_i A^{(1,0)}(\sigma_i,\sigma_i)}{3840 \sigma'_i{}^3}
-\frac{ \sigma''_i A^{(1,0)}(\sigma_i,\sigma_i)}{160 \sigma'_i{}^2}
+\frac{7 A(\sigma_i,\sigma_i) A^{(1,0)}(\sigma_i,\sigma_i)}{96 \sigma'_i{}^3}
+\frac{A^{(1,1)}(\sigma_i,\sigma_i)}{384 \sigma'_i{}^2}
+\frac{5 A^{(2,0)}(\sigma_i,\sigma_i)}{768 \sigma'_i{}^2}
\right)
$$
$$
+ \frac{1}{\varepsilon^2}
 \sum_{i=1}^{2s} \sum_{j \ne i}^{2s} \left(
 -\frac{1}{768 (\sigma_i-\sigma_j)^3 \sigma'_i}
+\frac{1}{48 (\sigma_i-\sigma_j)^3 \sigma'_j}
+\frac{\sigma'_j}{48 (\sigma_i-\sigma_j)^3 {\sigma'_i}^2}
\right.
$$
$$
\left.
-\frac{\sigma''_i}{384 (\sigma_i-\sigma_j)^2 {\sigma_i}^2}
+\frac{{\sigma''_i}}{10 (\sigma_i-\sigma_j)^2 \sigma'_i \sigma'_j}
+\frac{\sigma''_j}{1536 (\sigma_i-\sigma_j)^2 \sigma'_i \sigma'_j}
+ \frac{\sigma''_i \sigma''_j}{384 (\sigma_i-\sigma_j) \sigma'_i{}^2 \sigma'_j}
\right.
$$
$$
\left.
+\frac{A(\sigma_i,\sigma_i)}{128 (\sigma_i-\sigma_j)^2 \sigma'_i{}^2}
-\frac{A(\sigma_i,\sigma_i)}{3 (\sigma_i-\sigma_j)^2 \sigma'_i \sigma'_j}
-\frac{\sigma''_j A(\sigma_i,\sigma_i)}{128 (\sigma_i-\sigma_j) \sigma'_i{}^2 \sigma'_j}
\right.
$$
$$
\left.
+\frac{A(\sigma_i,\sigma_j)}{24 (\sigma_i-\sigma_j)^2 \sigma'_i{}^2}
+\frac{A(\sigma_i,\sigma_j)}{48 (\sigma_i-\sigma_j)^2 \sigma'_j{}^2}
-\frac{A(\sigma_i,\sigma_j)}{48 (\sigma_i-\sigma_j)^2 \sigma'_i \sigma'_j}
+ \frac{\sigma''_i \sigma''_j A(\sigma_i,\sigma_j)}{512 \sigma'_i{}^2 \sigma'_j{}^2}
+\frac{\sigma''_i {\sigma''_j} A(\sigma_i,\sigma_j)}{1536 {\sigma'_i}^3 {\sigma'_j}}
\right.
$$
$$
\left.
-\frac{{\sigma''_j}A(\sigma_i,\sigma_i) A(\sigma_i,\sigma_j)}{192 {\sigma'_i}^2 {\sigma'_j}^2}
-\frac{{\sigma''_j} A(\sigma_i,\sigma_i) A(\sigma_i,\sigma_j)}{384 {\sigma'_i}^3 {\sigma'_j}}
-\frac{A(\sigma_i,\sigma_j)^2}{48 (\sigma_i-\sigma_j) {\sigma'_i} {\sigma'_j}^2}
+\frac{A(\sigma_i,\sigma_j)^2}{24 (\sigma_i-\sigma_j) {\sigma'_i}^2 {\sigma'_j}}
+\frac{A(\sigma_i,\sigma_j)^3}{48 {\sigma'_i}^2 {\sigma'_j}^2}
\right.
$$
$$
\left.
-\frac{A(\sigma_j,\sigma_j)}{384 (\sigma_i-\sigma_j)^2 {\sigma'_i} {\sigma'_j}}
-\frac{{\sigma''_i} A(\sigma_j,\sigma_j)}{96 (\sigma_i-\sigma_j) {\sigma'_i}^2 {\sigma'_j}}
+\frac{A(\sigma_i,\sigma_i)A(\sigma_j,\sigma_j)}{32 (\sigma_i-\sigma_j) {\sigma'_i}^2 {\sigma'_j}}
-\frac{{\sigma''_i} A(\sigma_i,\sigma_j) A(\sigma_j,\sigma_j)}{128 {\sigma'_i}^2 {\sigma'_j}^2}
\right.
$$
$$
\left.
-\frac{{\sigma''_i} A(\sigma_i,\sigma_j) A(\sigma_j,\sigma_j)}{384 {\sigma'_i}^3{\sigma'_j}}
+\frac{A(\sigma_i,\sigma_i) A(\sigma_i,\sigma_j) A(\sigma_j,\sigma_j)}{48 {\sigma'_i}^2 {\sigma'_j}^2}
+\frac{A(\sigma_i,\sigma_i) A(\sigma_i,\sigma_j) A(\sigma_j,\sigma_j)}{96 {\sigma'_i}^3 {\sigma'_j}}
\right.
$$
\EQ{
\left.
-\frac{{\sigma''_i} {\cal L}^{(2)}_j(\sigma_i)}{512{\sigma'_i}^2}
+\frac{A(\sigma_i,\sigma_i) {\cal L}^{(2)}_j(\sigma_i)}{192 {\sigma'_i}^2}
+\frac{{\cal L}^{(2)}_j{}'(\sigma_i)}{1536 {\sigma'_i}}
-\frac{{\sigma''_j} A^{(0,1)}(\sigma_i,\sigma_j)}{1536 {\sigma'_i}{\sigma'_j}^2}
+\frac{A(\sigma_j,\sigma_j) A^{(0,1)}(\sigma_i,\sigma_j)}{384 {\sigma'_i} {\sigma'_j}^2}
\right)
\label{F2}
}
Let us consider the limit
\EQ{
y^2 = \varepsilon^2 \sigma_{2s}(x)  \longrightarrow 
\varepsilon^2 \left( x^s - a^s \right) \ , \qquad
a \rightarrow 0
}
where $s$ of the branch points come together.
In the double-scaling limit
\EQ{
a \rightarrow 0 \ , \quad
\varepsilon \rightarrow \infty \ , \quad
\Delta = \varepsilon a^{s/2+1} = cnst
}
we find
\EQ{
F_2 \ \longrightarrow \ \frac{F_g(\Sigma_-)}{\Delta^2} \ ,
\qquad \Sigma_- : y^2_- = \tilde{x}^s - \tilde{a}^s
}
as explained in Section \ref{DoubleFg}.  However,
the final result will not  simplify in general. 
In fact, it depends on the details of the near-critical spectral curve,
which has genus $[(s-1)/2]$.
An exception to this is given by the case of the conifold singularity,
where the original spectral curve becomes in the limit
\EQ{
y^2 \approx \varepsilon^2 (x-a)(x-b) \ , \qquad a,b \rightarrow 0 \ ,
}
which is essentially the spectral curve associated to a Gaussian matrix model \cite{TLG}.
As in Section \ref{DoubleFg}, it is convenient to choose one of the A-cycles
of the original spectral curve to be a loop encircling the cut going from 
branch point $a$ to branch point $b$. This cycle will reduce to an A-cycle
on the near-critical spectral curve
\EQ{
y^2_- = \tilde{\sigma}(\tilde{x}) = (\tilde{x} - \tilde{a})(\tilde{x} - \tilde{b})  
}
where, as before, the tilded quantities are finite in the limit $a,b \rightarrow 0$.
The above near-critical curve is actually a Riemann sphere.
In particular, one can check by evaluating the residues of all the integrands
at infinity that all periods of the form
\EQ{
\int_A \frac{d \tilde{x}}{\sqrt{\tilde{\sigma}(\tilde{x})}} \frac{1}{(\tilde{x}-\tilde{\sigma}_i)^n}
}
are zero. The expression of $F_2$ simplifies dramatically
\EQ{
F_2 \rightarrow - \frac{4}{15 \varepsilon^2 (a-b)^4} 
= - \frac{1}{240 S^2} \  ,
\label{F2conifoldcheck}}
where
\EQ{
S = \int_A y \, dx  \rightarrow \frac{\varepsilon(a-b)^2}{8} \sim \Delta \ ,
}
and in terms of $S$ the genus zero free energy is given by
\EQ{
F_0 \ \approx \   \frac{1}{2} S \frac{\partial F_0}{\partial S} \  \approx \ \frac{1}{2} S^2 \log S \ .
}
Thus, \eqref{F2conifoldcheck} indeed matches the expected result 
for the genus two free energy at a conifold singularity \cite{BCOV},
which is equivalent to the $c=1$ non-critical bosonic string \cite{GV}.
This particular singularity is obtained from a $2$-cut solution with a 
cubic superpotential in the limit where the two cuts touch each other. 
The fact that this limit should be equivalent to the $c=1$ non-critical string  
was also observed in \cite{DGKV}.

\section{Conclusion}

The class of matrix model DSLs that
are associated to the large $N$ field theory DSLs introduced in
\cite{BD} define a class of $c \leq 1$ non-critical bosonic strings \cite{TLG}.
They fall into different universality classes from the ones
usually considered in the old matrix model.  
We argued that these non-critical bosonic strings are related 
to the topological twist of non-critical superstring backgrounds
of the form $SL(2)/U(1) \times LG(X^n)$ that are dual to the
large $N$ double-scaled field theory and the associated
four-dimensional double-scaled LST at the corresponding $A_{n-1}$
singularity. 
To study the matrix models, and the relevant multicut solutions, 
we used the techniques of Chekhov and Eynard based on loop equations. 
These allow to show in general that the scaling
of the higher genus terms in the perturbative expansion of the
matrix model free energy matches precisely the scaling of 
the topological B model free energy in the vicinity of the Calabi-Yau
singularity, which is consistent with the Dijkgraaf-Vafa correspondence.  
We also evaluated the genus one and two terms explicitly
for the $A_{n-1}$ singularities, recovering the conifold result in the $n=2$ case. 
These techniques allow to study multicut solutions 
where the "old matrix model" tools are not generally available,
but further work would be needed to find the exact expression of the perturbative
matrix model free energy at all orders for the $A_{n-1}$ singularities with
$n > 2$. In particular, it would be interesting to see if this perturbative 
series needs a non-perturbative completion like in the conifold case. 
Such completion should correspond to $D$-brane effects on the non-critical 
string side as in \cite{AKK}.
It would also be interesting to perform the topological twist of the 
$SL(2)/U(1) \times LG$ model and determine the non-critical bosonic 
string explicitly.

\noindent
{\bf Acknowledgments.} I would like to thank Nick Dorey, Tim Hollowood, Luis Miramontes, Sameer Murthy and Asad Naqvi for various discussions and comments.

\startappendix

\Appendix{Some double-scaling formulae}\label{AppA}

In this appendix, we consider the double-scaling limit of various
quantities defined on the curve $\Sigma$ \eqref{mc}. This is most
conveniently done in the basis $\{\tilde A_i,\tilde B_i\}$ of 1-cycles
described in Section \ref{DoubleFg}. In particular, for $i\leq[n/2]$ these are
cycles on the near-critical curve $\Sigma_-$ in the double-scaling
limit.   

The key quantities that we will need are the periods
\EQ{
M_{ij} = \oint_{\tilde B_j} 
\frac{x^{i-1}}{\sqrt{\sigma(x)}}\, dx 
\ ,
\qquad
N_{ij} =
\oint_{\tilde A_j} \frac{x^{i-1}}{\sqrt{\sigma(x)}}\, dx \ .
\label{MijNijapp}
}
First of all, let us focus on $N_{ij}$ where $j\leq[n/2]$, but $i$ arbitrary.
By a simple scaling argument, as $a\to0$,
\EQ{
N_{ij} 
=
\int_{b_{(j)}^-}^{b_{(j)}^+} \frac{x^{i-1}}{\sqrt{B(x)}}\, dx
\longrightarrow
a^{i-n/2}
\int_{\tilde b_{(j)}^-}^{\tilde b_{(j)}^+} 
\frac{\tilde x^{i-1}}{\sqrt{ \tilde B(\tilde{x}) }}\, d \tilde x  
= a^{i-n/2}
\, f_{ij}^{(N)} ( \tilde{b}_l) \ ,
\label{N--}
}
for some function $f_{ij}^{(N)}$ of the branch points of $\Sigma_-$.
Here, $b_{(j)}^\pm$ are the two branch points enclosed by the cycle
$\tilde A_j$. A similar argument shows that $M_{ij}$ scales  in the
same way:
\EQ{
M_{ij}\longrightarrow a^{i-n/2}
\, f_{ij}^{(M)} ( \tilde{b}_l) \ .
}
So both
$N_{ij}$ and $M_{ij}$, for $i,j,\leq[n/2]$, diverge in the limit 
$a \to 0$. On the contrary, by using a similar argument, 
it is not difficult to see that, for $j>[n/2]$, $N_{ij}$ and $M_{ij}$ 
are analytic as $a\to0$ since the integrals are over non-vanishing
cycles. 

In summary, in the limit $a \to 0$, the matrices
$N$ and $M$ will have the following block structure
\EQ{
N \longrightarrow \left(
\begin{array}{cc}
N_{--} & N_{-+}^{(0)} \\
0 & N^{(0)}_{++} \\
\end{array} 
\right)\ , 
\qquad
M \longrightarrow \left(
\begin{array}{cc}
M_{--} & M_{-+}^{(0)} \\
0 & M^{(0)}_{++} \\
\end{array} 
\right) \ ,
\label{lim}
}
where by $-$ or $+$ we denote indices in the 
ranges $\{1, \ldots, [n/2]\}$ and $\{[n/2]+1,\ldots,s-1\}$
respectively. In \eqref{lim}, $N_{--}$ and $M_{--}$ are divergent
while the remaining quantities are finite as $a\to0$.

We also need the inverse $L=N^{-1}$. 
In the text, we use the polynomials $L_j(x)=\sum_{k=1}^{s-1}L_{jk}x^{k-1}$,
which enter the expression of the 
holomorphic 1-forms associated to our basis of 1-cycles,
\EQ{
\oint_{\tilde A_i}\omega_j=\delta_{ij}\  .
}
These 1-forms are equal to
\EQ{
\omega_j(x) = \frac{L_j(x)}{\sqrt{\sigma(x)}}\, dx \ =  
\frac{\sum_{k=1}^{s-1} L_{jk} x^{k-1}}{\sqrt{\sigma(x)}}\, dx \ , \qquad
\oint_{A_i} \omega_j(x) = \delta_{ij}
\label{hold}
}
where $i,j=1,\ldots,s-1$. From the behaviour of $N$ in
the limit $a\to0$, we have
\begin{equation}
L = N^{-1}
\longrightarrow \left(
\begin{array}{cc}
N^{-1}_{--} & {\cal N} \\
0 & \left( N^{(0)}_{++} \right)^{-1} \\
\end{array} 
\right) \ ,
\qquad 
{\cal N} = - N_{--}^{-1} \, N_{-+}^{(0)} \, 
\left( N_{++}^{(0)} \right)^{-1} \ . 
\label{NinvLimit}\end{equation}
Since $N_{--}$ is singular we see that $L$ is block diagonal in the
limit $a\to0$. This is just an expression of the fact that the curve
factorizes $\Sigma\to\Sigma_-\cup\Sigma_+$ as $a\to0$. In this
limit, using the scaling of elements of $L_{jk}$, we
find, for $j\leq[n/2]$,
\EQ{
\omega_j\longrightarrow \frac{\sum_{k=1}^{[n/2]}(f^{(N)})^{-1}_{jk}\tilde
  x^{k-1}}{\sqrt{\tilde B(\tilde x)}}d\tilde x=\tilde\omega_j\ .
}
the holomorphic 1-forms of $\Sigma_-$. While for $j>[n/2]$, 
\EQ{
\omega_j\longrightarrow \frac{\sum_{k>[n/2]}^{s-1}(N_{++}^{(0)})^{-1}_{jk}
x^{k-n/2-1}}{\sqrt{F(x)}}dx\ ,
}
are the holomorphic 1-forms of $\Sigma_+$.

\Appendix{The explicit expression of $\chi^{(n)}_i(p)$}

Using the formalism developed in \cite{Eynard} and
\cite{ChekhovEynard} to solve the matrix model loop equations, one
can easily find the expression of the differentials
$\chi^{(n)}_i(p)$ defined by
\EQ{
\left( \hat{K} - 2 W_0(p) \right) \chi^{(n)}_i(p) =
\frac{1}{(p-\sigma_i)^n} \ ,
}
where $\sigma_i$ is a branch point of the matrix model spectral curve.
These $1$-differentials appear quite naturally in the expression of higher loop
correlators in the matrix model and in the integration steps leading to $F_2$.
We have
\EQ{
\chi^{(n)}_i(p) = Res_{q \to \sigma_i} \left( \frac{dS_i(p,q)}{2 y(q)} \,
\frac{1}{(q-\sigma_i)^n} \right) 
\label{chin}} 
Given the expression of $dS_i(p,q)$, we can easily perform a Taylor expansion
in $q$ around the branch point $\sigma_i$ and find the residue. 
We will mainly consider the case where the matrix model spectral curve
has no double points, setting 
\EQ{
y^2 = \varepsilon^2 \sigma_{2s}(x) \ .
}
In this case
$$
\frac{dS_i(p,q)}{y(q)} = \frac{dS_i(p,q)}{\varepsilon
\sqrt{\sigma(q)}} = \frac{1}{\varepsilon\sqrt{\sigma(p)}} \left(
\frac{1}{p-q} - \sum_{j=1}^{s-1} L_j(p) \int_{A_j}
\frac{dx}{\sqrt{\sigma(x)}} \frac{1}{(x-q)} \,\,  \right) dp
$$
Note also that the expression in brackets is analytic in $q$. Then,
for instance, we find that
$$
\chi^{(1)}_i(p)
= \frac{1}{2 \varepsilon\sqrt{\sigma(p)}} \left(
\frac{1}{p-\sigma_i} - \sum_{j=1}^{s-1} L_j(p) \int_{A_j}
\frac{dx}{\sqrt{\sigma(x)}} \frac{1}{(x-\sigma_i)} \,\,  \right) dp
$$
\begin{equation}
=
\frac{1}{2 \varepsilon\sqrt{\sigma(p)}} \left(
\frac{1}{p-\sigma_i}
+ {\cal L}_i(p)
\right) dp
=
\frac{1}{2 \varepsilon\sqrt{\sigma(p)}} \left(
\frac{1}{p-\sigma_i}
+ \frac{A(p,\sigma_i)}{\sigma'_i}
\right) dp
\label{chi1i}\end{equation}

\begin{equation}
\chi^{(2)}_i(p) = \frac{1}{2 \varepsilon\sqrt{\sigma(p)}} \left(
\frac{1}{(p-\sigma_i)^2} - \sum_{j=1}^{s-1} L_j(p) \int_{A_j}
\frac{dx}{\sqrt{\sigma(x)}} \frac{1}{(x-\sigma_i)^2} \,\, \right) dp
\label{chi2i}\end{equation}
and in general
\begin{equation}
\chi^{(n)}_i(p) = \frac{1}{2 \varepsilon\sqrt{\sigma(p)}}
\frac{1}{(n-1)!} \frac{d^{n-1}}{d q^{n-1}} \left( \frac{1}{p-q} -
\sum_{j=1}^{s-1} L_j(p) \int_{A_j} \frac{dx}{\sqrt{\sigma(x)}}
\frac{1}{(x-q)} \,\,  \right)|_{q=\sigma_i} dp
\label{chini}\end{equation}
The above expressions can be generalized to the case where
the spectral curve is of the form
\EQ{
y^2 = M(x)^2 \sigma(x) \ ,
}
\begin{equation}
\chi^{(n)}_i(p) = \frac{1}{2 \sqrt{\sigma(p)}}
\frac{1}{(n-1)!} \frac{d^{n-1}}{d q^{n-1}} \frac{1}{M(q)} \left( \frac{1}{p-q} -
\sum_{j=1}^{s-1} L_j(p) \int_{A_j} \frac{dx}{\sqrt{\sigma(x)}}
\frac{1}{(x-q)} \,\,  \right)|_{q=\sigma_i} dp \ .
\label{chiniM}\end{equation}

\Appendix{Evaluation of $W^{(1)}(p)$ and $W(p,p,q)$}

In this section, we are going to evaluate two loop-functions whose
expression is needed later for $F_2$, the genus one one-loop function
$W^{(1)}(p)$ and the genus zero three-loop function $W(p,p,q)$.
Let us start from $W^{(1)}(p)$. Using the diagrammatic rules of
\cite{Eynard} we find
$$
W^{(1)}(x_2) = \sum_{i=1}^{2s} Res_{x_1 \to \sigma_i} \left(
\frac{dS_i(x_2,x_1)}{2 y(x_1)}\, W(x_1,x_1) \right)
$$
$$
= \sum_{i=1}^{2s} Res_{x_1 \to \sigma_i} \left[ \frac{dS_i(x_2,x_1)}{2
y(x_1)}\, \left(
\frac{1}{16(x_1-\sigma_i)^2}  + 
\frac{B_i}{x_1-\sigma_i} \right) \right]
$$
\begin{equation}
=
\sum_{i=1}^{2s} \frac{1}{16} \chi^{(2)}_i(x_2) +
B_i \chi^{(1)}_i(x_2) \ ,
\label{W(1)}\end{equation}
where
\EQ{
B_i \equiv \left(
-\frac{\sigma''(\sigma_i)}{8 \sigma'(\sigma_i)} + \sum_{j\ne i}
\frac{1}{8(\sigma_i-\sigma_j)} + \frac{A(\sigma_i,\sigma_i)}{4 \sigma'(\sigma_i)}
\right)
= \left(
-\frac{\sigma''(\sigma_i)}{16 \sigma'(\sigma_i)} + \frac{A(\sigma_i,\sigma_i)}{4 \sigma'(\sigma_i)}
\right)
\ .
\label{Bi}}
This is exactly the expression given for instance in \cite{ChekhovG=1},
once we use the identity
$$
A(\sigma_i,\sigma_i) = {\cal L}_i(\sigma_i) \,\sigma'(\sigma_i) \ .
$$
Then, let us evaluate the genus zero $3$-loop function with two
coincident arguments $W(x_2,x_2,x_3)$. We find
$$
W(x_2,x_2,x_3) 
= \sum_{i=1}^{2s} Res_{x_1 \to \sigma_i}
\left( \frac{dS_i(x_2,x_1)}{2 y(x_1)}\, W(x_1,x_2) W(x_1,x_3)
\right) 
$$
$$
= \sum_{i=1}^{2s} Res_{x_1 \to \sigma_i} \left[ \frac{dS_i(x_2,x_1)}{2
y(x_1)}\, \left( \frac{\sigma'(\sigma_i)}{4(x_2-\sigma_i)\sqrt{\sigma(x_2)}} +
\frac{A(x_2,\sigma_i)}{4 \sqrt{\sigma(x_2)}} \right) \right.
$$
$$
\left. \left( \frac{\sigma'(\sigma_i)}{4(x_3-\sigma_i)\sqrt{\sigma(x_3)}} +
\frac{A(x_3,\sigma_i)}{4 \sqrt{\sigma(x_3)}} \right)
\frac{1}{\sigma'(\sigma_i)(x_1-\sigma_i)} \right] 
$$
$$
= \sum_{i=1}^{2s} \left(
\frac{\sigma'(\sigma_i)}{4(x_2-\sigma_i)\sqrt{\sigma(x_2)}} +
\frac{A(x_2,\sigma_i)}{4 \sqrt{\sigma(x_2)}} \right) \left(
\frac{\sigma'(\sigma_i)}{4(x_3-\sigma_i)\sqrt{\sigma(x_3)}} +
\frac{A(x_3,\sigma_i)}{4 \sqrt{\sigma(x_3)}} \right)
\frac{\chi^{(1)}_i(x_2)
}{\sigma'(\sigma_i)}
$$
\begin{equation}
= \sum_{i=1}^{2s} \frac{\varepsilon^2 \sigma'(\sigma_i)}{4}
\chi^{(1)}_i(x_2){}^2 \chi^{(1)}_i(x_3) \  .
\label{W3}\end{equation}


\begin{thebibliography}{99}

\bibitem{BD}
  G.~Bertoldi and N.~Dorey,
  ``Non-critical superstrings from four-dimensional gauge theory,''
  JHEP {\bf 0511} (2005) 001
  [arXiv:hep-th/0507075].

\bibitem{Kutasov:1990ua}
  D.~Kutasov and N.~Seiberg,
  ``Noncritical Superstrings,''
  Phys.\ Lett.\ B {\bf 251} (1990) 67.

\bibitem{Dijkgraaf:2002fc}
  R.~Dijkgraaf and C.~Vafa,
  ``Matrix models, topological strings, and supersymmetric gauge theories,''
  Nucl.\ Phys.\ B {\bf 644} (2002) 3
  [arXiv:hep-th/0206255].

\bibitem{DV3}
  R.~Dijkgraaf and C.~Vafa,
  ``On geometry and matrix models,''
  Nucl.\ Phys.\ B {\bf 644} (2002) 21
  [arXiv:hep-th/0207106].

\bibitem{DVPW}
  R.~Dijkgraaf and C.~Vafa,
 ``A perturbative window into non-perturbative physics,''
  arXiv:hep-th/0208048.

\bibitem{TLG}
  G.~Bertoldi, T.~J.~Hollowood and J.~L.~Miramontes,
  ``Double Scaling Limits in Gauge Theories and Matrix Models,''
  arXiv:hep-th/0603122.

\bibitem{Vafaconifold}
  C.~Vafa,
  ``A Stringy test of the fate of the conifold,''
  Nucl.\ Phys.\ B {\bf 447} (1995) 252
  [arXiv:hep-th/9505023].
 
\bibitem{AGNT}
  I.~Antoniadis, E.~Gava, K.~S.~Narain and T.~R.~Taylor,
  ``Topological amplitudes in string theory,''
  Nucl.\ Phys.\ B {\bf 413} (1994) 162
  [arXiv:hep-th/9307158].

\bibitem{BCOV}
  M.~Bershadsky, S.~Cecotti, H.~Ooguri and C.~Vafa,
  ``Kodaira-Spencer theory of gravity and exact results for quantum string
  amplitudes,''
  Commun.\ Math.\ Phys.\  {\bf 165} (1994) 311
  [arXiv:hep-th/9309140].

\bibitem{DiFrancesco:1993nw}
  P.~Di Francesco, P.~H.~Ginsparg and J.~Zinn-Justin,
  ``2-D Gravity and random matrices,''
  Phys.\ Rept.\  {\bf 254} (1995) 1
  [arXiv:hep-th/9306153].
  P.~H.~Ginsparg and G.~W.~Moore,
  ``Lectures on 2-D gravity and 2-D string theory,''
  arXiv:hep-th/9304011.
 
\bibitem{Dijkgraaf:2003xk}
  R.~Dijkgraaf and C.~Vafa,
  ``N = 1 supersymmetry, deconstruction, and bosonic gauge theories,''
  arXiv:hep-th/0302011.
   
\bibitem{ADKMV}
  M.~Aganagic, R.~Dijkgraaf, A.~Klemm, M.~Marino and C.~Vafa,
  ``Topological strings and integrable hierarchies,''
  Commun.\ Math.\ Phys.\  {\bf 261} (2006) 451
  [arXiv:hep-th/0312085].
      
\bibitem{Eynard}
  B.~Eynard,
  ``Topological expansion for the 1-hermitian matrix model correlation
  functions,''
  JHEP {\bf 0411} (2004) 031
  [arXiv:hep-th/0407261].
  
\bibitem{ChekhovEynard}
  L.~Chekhov and B.~Eynard,
  ``Hermitean matrix model free energy: Feynman graph technique for all
  genera,''
  arXiv:hep-th/0504116.

\bibitem{GK}
A.~Giveon and D.~Kutasov,
``Little string theory in a double-scaling limit,''
JHEP {\bf 9910}, 034 (1999)
[arXiv:hep-th/9909110]. \\
A.~Giveon and D.~Kutasov,
``Comments on double scaled little string theory,''
JHEP {\bf 0001}, 023 (2000)
[arXiv:hep-th/9911039].

\bibitem{GKP}
A.~Giveon, D.~Kutasov and O.~Pelc,
``Holography for non-critical superstrings'',
JHEP {\bf 9910} (1999) 035
[arXiv:hep-th/9907178].

\bibitem{LST}
N.~Seiberg,
``New theories in six dimensions and matrix description of M-theory on
T**5 and T**5/Z(2),''
Phys.\ Lett.\ B {\bf 408} (1997) 98
[arXiv:hep-th/9705221]. \\
M.~Berkooz, M.~Rozali and N.~Seiberg,
``On transverse fivebranes in M(atrix) theory on T**5,''
Phys.\ Lett.\ B {\bf 408} (1997) 105
[arXiv:hep-th/9704089].

\bibitem{holog}
O.~Aharony, M.~Berkooz, D.~Kutasov and N.~Seiberg,
``Linear dilatons, NS5-branes and holography,''
JHEP {\bf 9810} (1998) 004
[arXiv:hep-th/9808149].

\bibitem{OV}
  H.~Ooguri and C.~Vafa,
  ``Two-Dimensional Black Hole and Singularities of CY Manifolds,''
  Nucl.\ Phys.\ B {\bf 463} (1996) 55
  [arXiv:hep-th/9511164].

\bibitem{GV}
  D.~Ghoshal and C.~Vafa,
  ``C = 1 string as the topological theory of the conifold,''
  Nucl.\ Phys.\ B {\bf 453} (1995) 121
  [arXiv:hep-th/9506122].

\bibitem{MV}
  S.~Mukhi and C.~Vafa,
  ``Two-dimensional black hole as a topological coset model of c = 1 string
  theory,''
  Nucl.\ Phys.\ B {\bf 407} (1993) 667
  [arXiv:hep-th/9301083].

\bibitem{Ashok:2005xc}
  S.~K.~Ashok, S.~Murthy and J.~Troost,
  ``Topological cigar and the c = 1 string: Open and closed,''
  arXiv:hep-th/0511239.

\bibitem{DGKV}
  R.~Dijkgraaf, S.~Gukov, V.~A.~Kazakov and C.~Vafa,
  ``Perturbative analysis of gauged matrix models,''
  Phys.\ Rev.\ D {\bf 68} (2003) 045007
  [arXiv:hep-th/0210238].
 
 \bibitem{Hori:2001ax}
  K.~Hori and A.~Kapustin,
  ``Duality of the fermionic 2d black hole and N = 2 Liouville theory as
  mirror symmetry,''
  JHEP {\bf 0108} (2001) 045
  [arXiv:hep-th/0104202].

\bibitem{Tong:2003ik}
  D.~Tong,
  ``Mirror mirror on the wall: On two-dimensional black holes and Liouville theory,''
  JHEP {\bf 0304} (2003) 031
  [arXiv:hep-th/0303151].
 
\bibitem{Eguchi:2004ik}
  T.~Eguchi and Y.~Sugawara,
  ``Conifold type singularities, N = 2 Liouville and SL(2,R)/U(1) theories,''
  JHEP {\bf 0501} (2005) 027
  [arXiv:hep-th/0411041].
 
\bibitem{Eguchi:2004yi}
  T.~Eguchi and Y.~Sugawara,
  ``SL(2,R)/U(1) supercoset and elliptic genera of non-compact Calabi-Yau
  manifolds,''
  JHEP {\bf 0405} (2004) 014
  [arXiv:hep-th/0403193].
 
\bibitem{Israel:2004jt}
  D.~Israel, A.~Pakman and J.~Troost,
  ``D-branes in N = 2 Liouville theory and its mirror,''
  Nucl.\ Phys.\ B {\bf 710} (2005) 529
  [arXiv:hep-th/0405259].
 
\bibitem{Rastelli:2005ph}
  L.~Rastelli and M.~Wijnholt, ``Minimal AdS(3),''
  arXiv:hep-th/0507037.
 
 \bibitem{Sahakyan:2005dh}
  D.~A.~Sahakyan and T.~Takayanagi,
  ``On the connection between N = 2 minimal string and (1,n) bosonic minimal
  string,''
  arXiv:hep-th/0512112.

\bibitem{Niarchos:2005ny}
  V.~Niarchos,
  ``On minimal N = 4 topological strings and the (1,k) minimal bosonic
  string,''
  arXiv:hep-th/0512222.

\bibitem{NN}
  S.~Nakamura and V.~Niarchos,
  ``Notes on the S-matrix of bosonic and topological non-critical strings,''
  JHEP {\bf 0510} (2005) 025
  [arXiv:hep-th/0507252].

\bibitem{Takayanagi0507}
  T.~Takayanagi,
  ``Notes on S-matrix of non-critical N = 2 string,''
  JHEP {\bf 0509} (2005) 001
  [arXiv:hep-th/0507065].

\bibitem{Takayanagi0503}
  T.~Takayanagi,
  ``$c < 1$ string from two dimensional black holes,''
  JHEP {\bf 0507} (2005) 050
  [arXiv:hep-th/0503237].
 
\bibitem{CSW}
  F.~Cachazo, N.~Seiberg and E.~Witten,
  ``Phases of N = 1 supersymmetric gauge theories and matrices,''
  JHEP {\bf 0302}, 042 (2003)
  [arXiv:hep-th/0301006].

\bibitem{CDSW}
  F.~Cachazo, M.~R.~Douglas, N.~Seiberg and E.~Witten,
  ``Chiral rings and anomalies in supersymmetric gauge theory,''
  JHEP {\bf 0212}, 071 (2002)
  [arXiv:hep-th/0211170].  
  
\bibitem{VY}
  G.~Veneziano and S.~Yankielowicz,
  ``An Effective Lagrangian For The Pure N=1 Supersymmetric Yang-Mills
  Theory,''
  Phys.\ Lett.\ B {\bf 113}, 231 (1982).

\bibitem{Eguchi:2003wv}
T.~Eguchi and Y.~Sugawara,
``Branches of N = 1 vacua and Argyres-Douglas points,''
JHEP {\bf 0305} (2003) 063
[arXiv:hep-th/0305050].

\bibitem{bert}
  G.~Bertoldi,
  ``Matrix models, Argyres-Douglas singularities and double-scaling limits,''
  JHEP {\bf 0306}, 027 (2003)
  [arXiv:hep-th/0305058].
 
\bibitem{ACKM}
  J.~Ambjorn, L.~Chekhov, C.~F.~Kristjansen and Y.~Makeenko,
  ``Matrix model calculations beyond the spherical limit,''
  Nucl.\ Phys.\ B {\bf 404} (1993) 127
  [Erratum-ibid.\ B {\bf 449} (1995) 681]
  [arXiv:hep-th/9302014].

\bibitem{Akemann}
  G.~Akemann,
  ``Higher genus correlators for the Hermitian matrix model with multiple
  cuts,''
  Nucl.\ Phys.\ B {\bf 482} (1996) 403
  [arXiv:hep-th/9606004].

\bibitem{AmbjornAkemann}
  J.~Ambjorn and G.~Akemann,
  ``New universal spectral correlators,''
  J.\ Phys.\ A {\bf 29} (1996) L555
  [arXiv:cond-mat/9606129]. 

\bibitem{Makeenko}
  Y.~Makeenko,
  ``Loop equations in matrix models and in 2-D quantum gravity,''
  Mod.\ Phys.\ Lett.\ A {\bf 6} (1991) 1901.

\bibitem{ChekhovG=1}
  L.~Chekhov,
  ``Genus one correlation to multi-cut matrix model solutions,''
  Theor.\ Math.\ Phys.\  {\bf 141} (2004) 1640
  [Teor.\ Mat.\ Fiz.\  {\bf 141} (2004) 358]
  [arXiv:hep-th/0401089].

\bibitem{KMT}
  A.~Klemm, M.~Marino and S.~Theisen,
  ``Gravitational corrections in supersymmetric gauge theory and matrix
  models,''
  JHEP {\bf 0303} (2003) 051
  [arXiv:hep-th/0211216].

\bibitem{DST}
  R.~Dijkgraaf, A.~Sinkovics and M.~Temurhan,
  ``Matrix models and gravitational corrections,''
  Adv.\ Theor.\ Math.\ Phys.\  {\bf 7} (2004) 1155
  [arXiv:hep-th/0211241].

\bibitem{KostovCFT}
  I.~K.~Kostov,
  ``Conformal field theory techniques in random matrix models,''
  arXiv:hep-th/9907060.

\bibitem{Moore}
  G.~W.~Moore,
  ``Matrix Models Of 2-D Gravity And Isomonodromic Deformation,''
  Prog.\ Theor.\ Phys.\ Suppl.\  {\bf 102} (1990) 255.
  
\bibitem{Vasiliev}
  D.~Vasiliev,
  ``Determinant formulas for matrix model free energy,''
  arXiv:hep-th/0506155.

\bibitem{AD}
  P.~C.~Argyres and M.~R.~Douglas,
``New phenomena in SU(3) supersymmetric gauge theory,''
  Nucl.\ Phys.\ B {\bf 448}, 93 (1995)
  [arXiv:hep-th/9505062].

\bibitem{ARSW}
P.~C.~Argyres, M.~Ronen Plesser, N.~Seiberg and E.~Witten,
 ``New N=2 Superconformal Field Theories in Four Dimensions'',
Nucl.\ Phys.\ B {\bf 461} (1996) 71
[arXiv:hep-th/9511154].

\bibitem{EHIY}
T.~Eguchi, K.~Hori, K.~Ito and S.~K.~Yang,
``Study of $N=2$ Superconformal Field Theories in $4$ Dimensions,''
Nucl.\ Phys.\ B {\bf 471} (1996) 430
[arXiv:hep-th/9603002].

\bibitem{AKK}
  S.~Y.~Alexandrov, V.~A.~Kazakov and D.~Kutasov,
  ``Non-perturbative effects in matrix models and D-branes,''
  JHEP {\bf 0309} (2003) 057
  [arXiv:hep-th/0306177].

\end{thebibliography}
\end{document}